\documentclass[preprint,12pt,sort&compress]{elsarticle}
\usepackage[left=2.0cm,right=2.0cm,top=1cm,bottom=2cm,bindingoffset=0cm]{geometry}
\usepackage{natbib}
\usepackage{amsmath,amssymb,amsthm}
\usepackage{multicol}

\usepackage[config, labelfont={sf,bf}, textfont=sf]{subfig}
\graphicspath{{./}}
\usepackage[suffix=, outdir=./]{epstopdf}

\usepackage{enumitem}

\usepackage[T2A]{fontenc}
\usepackage[utf8]{inputenc}
\usepackage[english]{babel}

\usepackage{array}
\usepackage{multirow}
\usepackage{hhline}
\usepackage{makecell}

\usepackage[usenames,dvipsnames]{xcolor}

\usepackage{hyperref}
\hypersetup{
  breaklinks=true,  
  colorlinks=true,
  urlcolor=urlcolor,
  linkcolor=linkcolor,
  citecolor=citecolor,
  }

\definecolor{urlcolor}{rgb}{0,.145,.698}
\definecolor{linkcolor}{rgb}{.71,0.21,0.01}
\definecolor{citecolor}{rgb}{.12,.54,.11}

\usepackage{listings}

\definecolor{dkgreen}{rgb}{0,0.6,0}
\definecolor{gray}{rgb}{0.5,0.5,0.5}
\journal{Journal Name}

\begin{document}

\begin{frontmatter}

\title{Forecasting and control in overlapping generations model: \\ chaos stabilization via artificial intelligence}

\author[hse]{T.A. Alexeeva}
\author[vsb-tuo]{Q.B. Diep}
\author[spbu,ipm]{N.V. Kuznetsov\footnote{Corr. author email nkuznetsov239@gmail.com}}
\author[spbu]{T.N. Mokaev}
\author[vsb-tuo]{I. Zelinka}

\address[hse]{St.~Petersburg School of Physics, Mathematics and Computer Science, HSE University, \\
194100 St. Petersburg, Kantemirovskaya ul., 3, Russia}
\address[vsb-tuo]{Department of Computer Science, Faculty of Electrical Engineering and Computer Science, V\v{S}B-TUO, 17.listopadu 2172/15, 708 00 Ostrava-Poruba, Czech Republic}
\address[spbu]{Faculty of Mathematics and Mechanics,
St. Petersburg State University, 198504 Peterhof, St. Petersburg, Russia}
\address[ipm]{Institute for Problems in Mechanical Engineering RAS, 199178 St. Petersburg, V.O., Bolshoj pr., 61, Russia}

\begin{abstract}

Irregular, especially chaotic, behavior is often undesirable for economic processes because it presents challenges for predicting their dynamics. In this situation, control of such a process by its mathematical model can be used to suppress chaotic behavior and to transit the system from irregular to regular dynamics.

In this paper, we have constructed an overlapping generations model with a control function. By applying evolutionary algorithms we showed that in the absence of control, both regular and irregular behavior (periodic and chaotic) could be observed in this model. We then used the synthesis of control by the Pyragas control method with two control parameters to solve the problem of controlling the irregular behavior of the model. We solved a number of optimization problems applying evolutionary algorithms to select control parameters in order to ensure stability of periodic orbits. We compared qualitative and quantitative characteristics of the model’s dynamics before and after applying control and verified the results obtained using simulation.

We thus demonstrated that artificial intelligence technologies (in particular, evolutionary algorithms) combined with the Pyragas control method are well suited for in-depth analysis and stabilization of irregular dynamics in the model considered in this paper.
\end{abstract}


\end{frontmatter}

\section{Introduction} \label{S:1}

Currently the world economy develops in conditions of complexity, uncertainty and unpredictability related to technology transformation, changes in the economic structure, as well as a number of exogenous challenges, including climatic, energy, and epidemiological cataclysms \cite{KindlebergerA-book-2015,Scheffer_etal21-2012,Battiston_et.al.22-2016}.
These challenges stimulated posing of new problems both in the real economy and in economic science, which led to conceptual shifts in the testing of hypotheses about the functioning of economic systems, construction of the mathematical models as well as analysis and forecasting their dynamics \cite{DuffyMcN-2001-JEDC,EinavL-Sci-2014, JordanM-Sci-2015,Goodfellow-Book-2016,SendhilS-2017-JEP,AghionJJ-2019-UCP,AgrawalGG-2019-Book,Duarte-2018}.
Economists need to know how models can impact in the real world and they often focus not only on forecasts but also on model inference, on understanding and interpretation the model parameters. Nevertheless, economic and fiscal policies conceived by governments, central banks, and other decision-makers heavily depend on economic forecasts, in particular during times of economic, societal, and natural turmoils, policy makers must support their decisions by providing and communicating explanations for the action taken. 
Therefore, they are interested in the economic implications associated with model predictions. 
The success of mathematical modeling significantly depends on the ability to obtain rich information about economic activity and uncover complex economic relationships that could be useful to forecast the economy in normal time, and also to identify early signals of problems in markets before crises \cite{Consoli-2021-BookDataSciEc}.
In order to get a more complete picture of the state of the economy and its future behavior, researchers increasingly rely on both theoretical concepts and various types of economic data, including detailed microdata and big data. This allows studying behavior of complex, large scale, nonlinear, models with a large number of variables or parameters, as well as forecasting the dynamics of economic processes. 
Given such features of modern models, special technologies adapted to powerful computing resources are required to analyze them, forecast, and verify the predictions. In this regard, it is difficult to offer a more suitable tool than methods based on artificial intelligence (AI) technologies, since AI algorithms can handle severe nonlinearities, are easy to dynamically scale to large state spaces and thousands computer nodes \cite{Consoli-2021-BookDataSciEc,MaliarMW-Why-2019,FernandezNSLV-NBER-DL-2020,BeaudryGP-AER-2020,LeCunBH-2015-Nature,SandersFarmerG-SR-2018}. 
There is a vast literature \cite{Bloom-2014-JEP,BakerBD-2016-QJE,HansenMcM-2016-JIE,CasellaFVH-2020,Consoli-2021-BookDataSciEc,GiannoneRS-2008-JME,Marx-2013-Nature-BigD,BarbagliaCM-4-2020,ConsoliPT-14-ML-2020,Tetlock-37-JF-2007,ArmengolZ-19-JUE2005,ArmengoiZ-18-IER-2004,CetinaPR-23-JFS-2018,BoivinG-2006,MaliarMW-Why-2019,FernandezHN-NBER-2019,FernandezG-Est-2020,AzinovicGS-2019_SSRN,KudybaD-HarvardBR-2018,ParkersW-Sci-2015} that provides impressive examples of applying various AI technologies, such as machine learning (ML) methods, like support vector machines, decision trees, random forests; deep learning (DL), including reinforcement learning; semantic web technologies, involving natural language processing; and evolutionary algorithms (EAs) \cite{price2013differential,davendra2016self,Zelinka-EAAI-2009,Bilal-EAAI-2020,dasgupta2013evolutionary,back1993overview}, to solve a wide spectrum of the practice-focused and theoretical problems in the economy.
Nonlinear behavior of the modern models presents special challenges to the accuracy of forecasting, both in the short and medium to long run. 
If nonlinearities are present in the model, irregular dynamics and complex limit behavior could arise that may manifest themselves as unstable regimes or chaos. Chaotic behavior in a model of an economic system may leads to unpredictable events, complicate analytical and numerical study of the model, finding acceptable values of the model parameters, and hinder the accuracy of forecasts over longer horizons, thus undermining the predictive power of the model \cite{Grandmont-E1985,Adam-2003-RES-OLG}. This is undesirable from the point of view of policy makers aiming to stabilize aggregate fluctuations.
Numerous papers \cite{BenhabibD8-1981,Day2-1983,BarnettCh:J-1988,Medio6-1992,Hommes-1995,BrockHomEc-1997,Kopel-1997,BrockS14-1998,BarnettSert10-2000,Rosser:Book-2000,BenhabibSGU-AER-2002,WielandW-JEBO-2005,HommesBook12-2006,Neck:15-2009,SalariehAl:3-2009,AmritRA:13ARC-2011,CavalliNP:5-2017,Bella-CSF-2017,BellaMV-JET-2017,AlexeevaMP-2020-DUe,AlexeevaKMP-2021-JPCS,Barnett-2020,Galizia-QE-2021,Adam-2003-RES-OLG} attempted to explain such features of economic data as irregular and erratic microeconomic and macroeconomic fluctuations, financial and credit crises, structural changes, and overlapping waves of economic development from the point of view of chaos theory. A common theme in this literature is explaining the complexity and unpredictable behavior of economic processes by nonlinear dynamical models. 
Trajectories in such models, starting somewhere in the phase space, can be attracted not only to a stable stationary point or a periodic cycle, but to an irregular invariant set, including chaotic attractor \cite{LeonovK-2013-IJBC,KuznetsovL-2014-IFACWC,LeonovKM-2015-EPJST,Kuznetsov-2016}.
Additional complexity of the dynamics can be also associated with various unstable orbits embedded into the chaotic attractor of the dynamical system.
If the model that describes particular economic phenomenon exhibits such complex dynamics, its forecasting and control becomes a very important problem to be solved.
In this regard, the usage of AI technologies in combination with the classical control methods allows making significant progress in determining the qualitative properties of the model dynamics, including revealing of regular and irregular (periodic and chaotic) regimes, fine-tuning of the possible initial points, optimization of the model parameters, and stabilization of unstable orbits by using control procedures \cite{KuznetsovMKK-2020,AlexeevaBKM-IFAC-2020,AlexeevaKM-2021-CSF}.

In this paper, we demonstrate the effectiveness of applying the numerical and analytic approach grounded on EAs and the Pyragas control method \cite{Pyragas-1992,Pyragas-2006} to investigate and forecasting the irregular dynamics of macroeconomic processes using one of overlapping generations (OLG) models as an example. 
A pioneering OLG model, which was developed by Nobel laureates Paul Samuelson (1970) and Peter Diamond (2010) in \cite{Samuelson-1958-JPE,Diamond-1965}, is a representative of a very important class of low dimensional economic models with optimizing agents that are used to analyze the basic intertemporal choice of consumption and saving and the dynamic consequences of these choices, as well as to explore dynamics of education, retirement, capital accumulation, public policies, inflation, fiscal policies, etc. (see, e.g. \cite{Farmer-KW-2003-ET,BenhabibD-JEDC-1982,Barnett-1989-Book,DixonD-2013-Book,EvansH-2001-Book-OLG,Galor-1992-EcOLG,Arifovic-1995-JME-OLG,Woodford-1994-OLG,Adam-2003-RES-OLG,Weil-2008-OLG,Fehr-HB-2013-OLG-1,NishiyamaS-2014-OLG-2,ZodrowD-HB-2012-OLG-3,Kurz-HB-2009-OLG-4,Quadrini-HB-2015-OLG-5,Woodland-HB-2016-OLG-7,PeraltaAS-HB-2014-OLG-8,Brunnermeier-HB-2013-OLG-9,Adam-2003-RES-OLG,ArifovicHS-2019-OLG})
\footnote{
The reason interest in OLG models has not faded almost a century after \cite{Samuelson-1958-JPE} is well described in \cite{Weil-2008-OLG} dedicated to the 50th anniversary of publication of Samuelson’s paper: ``Like Mona Lisa’s enigmatic smile, the mysterious welfare properties of the overlapping generations model are, to a significant extent, responsible for its popularity—along with the many economic issues it has illuminated in the last half-century.''}.
Despite being low dimensional, OLG models are shown to exhibit a wide range of complex behaviors -- saddle-path converging toward the steady state \cite{Farmer-KW-2003-ET}, cycles \cite{BenhabibD-JEDC-1982}, divergence \cite{BenhabibD-JEDC-1982}, sunspot equilibria \cite{Woodford-1994-OLG}, multiple equilibria \cite{Adam-2003-RES-OLG}, and chaos \cite{BenhabibD-JEDC-1982}. 
In this paper, we derive a two period OLG model with production and endogenous labor choice which is represented by a discrete-time dynamical model arising from solutions of economic agents' dynamic optimization problems: for consumers and firms. 
The model is deterministic, where agents exist in an environment of perfect foresight and have exact information about the values of the model parameters and the trajectories of solutions in the phase space along which the dynamics of the model evolves. Even with parameter values that imply existence of a chaotic regime, starting from specific initial conditions could lead to switching to a trajectory with predictable dynamics by applying control, modeled as time varying government spending. In the absence of control, the model can behave both regularly and irregularly, including periodically and chaotically.
We show how such a model could be successfully studied by consecutive application of EAs and the Pyragas method. 
To this end, we chose the most powerful EA
 methods, i.e. differential evolution (DE) \cite{price2013differential} and the self-organized migration algorithm (SOMA) \cite{davendra2016self}.
To suppress the chaotic regime of the model's dynamics, we started by applying EAs to overcome a complex fractional-power nonlinearity of the model to find the suspected unstable periodic orbits embedded into the attractor. 
To refine these trajectories, we used the Pyragas method, then synthesized a time-delayed feedback control and found out its parameters by solving a particular optimization problem using EAs in such a way that the periodic trajectory became locally stable. 
Last but not least, using EAs and computational abilities of a supercomputer, we also solved an optimal control problem of maximizing the basin of attraction of the stabilized UPO, and fine-tuning of the possible initial points from which the state of the system is attracted to the specified trajectory.
Thereby, all the principal stages of the examination of the limiting dynamics of the OLG model were performed using the EAs. 
Our result shows that even for low-dimensional nonlinear models in the case of chaotic behavior, it is sometimes critical to use EAs combined with the computing power of supercomputers to be able to solve particular control and optimal control problems, as well as forecasting problems.

\section{The model} \label{S:2}

OLG models are a useful theoretical concept that allows one to construct economic theories, introduce and interpret various effects of economic policies, and understand how the economic system functions given the finite life cycle of economic agents.
First, these models allow us to explicitly consider life-cycle issues and demographic trends, including education and pension systems, and to study variety of problems associated with financing and reforming them, as well as to estimate the effect of population ageing on pension reforms and government fiscal policy \cite{Quadrini-HB-2015-OLG-5,Woodland-HB-2016-OLG-7}. Second, these models introduce a natural heterogeneity of agents belonging to different cohorts, which makes it possible to speak of intergenerational transfers (education as a transfer from employed population to the young, pensions as transfers from employed population to the elderly), to investigate mechanisms of trade between generations \cite{Farmer-KW-2003-ET}, and to estimate both short-run transitory and long-run dynamic macroeconomic effects of tax reforms \cite{NishiyamaS-2014-OLG-2,ZodrowD-HB-2012-OLG-3}. Third, these models naturally introduce incomplete markets as the agents do not have access to markets that existed before they were born and cannot trade in markets that open after their death. This creates an opportunity to deviate from the fundamental theorems of welfare economics \cite{Arrow-1963}, and can lead to such nontrivial economic phenomena with complex dynamics as sunspots, indeterminacy, over-accumulation of capital, etc (see, e.g. \cite{BenhabibNish-Book-Ch-2012,Woodford-AER-1987,Slobodyan-JEDC-2005}). The OLG models thus are of special interest for study, because they present a complicated environment with varied dynamics, including stationary states, cycles of all periods, and even chaotic dynamics \cite{ArifovicHS-2019-OLG}. 

Here we derive a new OLG model developing the ideas proposed in \cite{MendesM-2005}, who considered a two-period OLG model with two types of economic agents: firms and households (i.e. consumers). Economic agents live for two periods. In every period $t,$ there are two consumer cohorts (of size 1)\footnote{But we could also think that there is only one person in every generation.}: the one born at $t$ (young) and the one born at $t-1$ (old). 
Consumers maximize their welfare by solving a dynamic optimization problem subject to the budget constraint and determine work hours and savings. Consumers work only in period 1 but consume in both periods. 
In period $t,$ the young born at $t$ work, providing labor hours
$l_{t},$ and consume $c_{t}^{t}.$ Here a superscript $t$ denotes cohort, or time of birth, and 
a subscript is physical time. They consume a single good which also could be saved and turned into capital $k_{t}$, which will be used for production in the next period $t+1$. In the second period of their life, the agents cannot work, and could only consume their savings with interest.
 Profit maximizing firms use labor and capital to produce a single good used for consumption and investment. 
 In contrast to \cite{MendesM-2005}, our model takes full account of the old cohort's consumption. Additionally,  \cite{MendesM-2005} takes one of the fundamental model parameters -- $\gamma$ (labor elasticity), to be the control. This amounts to a new agent with different preferences appearing at each point in time when the controller changes $\gamma$. While agents' preferences could be stochastic, they cannot be controlled by any government or a social planner. In our model, we introduce government spending, which is an external non-fundamental variable. Government spending is financed by proportional labor tax, and is used as a control variable in a chaotic regime.

\subsection{Utility side}

Let $c_{t}=c_{t}^{t}$ is consumption of young at period $t$,
$c_{t+1}=c_{t+1}^{t}$ is consumption of young at period $t+1$ when they
will be old.
The only meaningful decision, therefore, is that of young agents, who at time
$t$ must make an optimal choice of their consumption while young ($c_{t}^{t}%
$), consumption when old ($c_{t+1}^{t}$), and working hours when young
($l_{t}$). The optimal behavior is represented in the optimization problem 
-- the {\it consumer problem}
\begin{equation}
\max \text{ }U_{t}   =u_{1}\left(  c_{t}^{t}\right)  +u_{2}\left(  c_{t+1}%
^{t}\right)  -v\left(  l_{t}\right)  ,
\label{cp:obj-function}
\end{equation}
\begin{equation}
\begin{aligned}
&s.t.\\
&c_{t}^{t}+k_{t}   =\left(  1-\tau_{t}\right)  w_{t}l_{t},\\
&c_{t+1}^{t}  =R_{t+1}k_{t},
\end{aligned}
\label{cp:constraints}
\end{equation}
where $u_{1}\left(  c_{t}^{t}\right)  $ is the utility of consuming while
young, $u_{2}\left(  c_{t+1}^{t}\right)  $ the (future) utility of consuming
when old, and $v\left(  l_{t}\right)  $ disutility of labor. In the first
period, the {\it budget constraint} \eqref{cp:constraints} says that consumption $c_{t}^{t}$ plus capital
($k_{t}$) equals the income of the young, given by their labor income
$w_{t}l_{t}$ net of proportional taxes (for example, profits tax) with rate $\tau_{t}$.
Here  $w_t$ is the real wage rate, $R_{t+1}$ is the gross real interest rate.
In the second period of their life (physical time $t+1$), the agents born at $t$ consume
their savings with interest. Non-negativity constraints form part of the model.

Plugging in both constraints into the utility function, we get the following
problem:%

\begin{equation}
\begin{aligned}
\max \text{ }\underset{k_{t},l_{t}}{U_{t}}=u_{1}\left(  \left(  1-\tau
_{t}\right)  w_{t}l_{t}-k_{t}\right)  +u_{2}\left(  R_{t+1}k_{t}\right)
-v\left(  l_{t}\right)  ,
\label{cp:max_obj_func}
\end{aligned}
\end{equation}
which produces the following first order conditions (FOCs):%
\begin{equation}
\begin{aligned}
&\frac{\partial U_{t}}{\partial k_{t}}  & =0:-u_{1}^{\prime}\left(  c_{t}%
^{t}\right)  +R_{t+1}u_{2}^{\prime}\left(  c_{t+1}^{t}\right)  =0,\\
&\frac{\partial U_{t}}{\partial l_{t}}  & =0:\left(  1-\tau_{t}\right)
w_{t}\cdot u_{1}^{\prime}\left(  c_{t}^{t}\right)  -v^{\prime}\left(
l_{t}\right)  =0.\end{aligned}
\label{cp:FOCs}
\end{equation}
From the second equation, we see that
\begin{equation}
\begin{aligned}
&\left(  1-\tau_{t}\right)
w_{t}=\frac{v^{\prime}\left(l_{t}\right)}{u_{1}^{\prime}\left(  c_{t}^{t}\right)}, \end{aligned}
\label{cp:FOC_wt}
\end{equation}
and from the first that
\begin{equation}
\begin{aligned}
 R_{t+1}=\frac{u_{1}^{\prime}\left(  c_{t}^{t}\right)}
{u_{2}^{\prime}\left(  c_{t+1}^{t}\right)}  . \end{aligned}
\label{cp:FOC_Rt}
\end{equation}
 Plugging both into the budget constraint \eqref{cp:constraints}, we get%
\begin{equation*}
\begin{aligned}
&c_{t}^{t}+k_{t}   =c_{t}^{t}+\frac{c_{t+1}^{t}}{R_{t+1}}=c_{t}^{t}+c_{t+1}^{t}%
\frac{u_{1}^{\prime}\left(  c_{t}^{t}\right)}{u_{2}^{\prime}\left(  c_{t+1}^{t}\right)}  =\\
& =\left(  1-\tau_{t}\right)  w_{t}l_{t}=\frac{v^{\prime}\left(  l_{t}\right)
l_{t}}{u_{1}^{\prime}\left(  c_{t}^{t}\right)}  ,\end{aligned}
\end{equation*}
\begin{equation}
c_{t}^{t}u_{1}^{\prime}\left(  c_{t}^{t}\right)  +c_{t+1}^{t}
u_{2}^{\prime}\left(  c_{t+1}^{t}\right)  -v^{\prime}\left(  l_{t}\right)
l_{t} = 0,
\label{cp:fin_budget_constraint_Euler}
\end{equation}
Using the same functional forms as in \cite{MendesM-2005}

\begin{equation}
\begin{aligned}
&u_1(c_t^t) = \frac{1}{\theta} \left( c_t^t \right)^{\theta} , \quad 0 < \theta < 1 \\
&u_2(c_{t+1}^{t}) = \frac{1}{\alpha} \left( c_{t+1}^{t} \right)^{\alpha} , \quad 0 < \alpha < 1\\
&v(c_t^t) = \frac{1}{\gamma} \left( c_t^t \right)^{\gamma} , \quad \gamma > 1 , \end{aligned}
\label{utility-1}
\end{equation}
we get the first equation of our model
\begin{equation}
\begin{aligned}
c_{t+1}^{t}=\left( l_{t}^{\gamma} - \left(  c_{t}^{t}\right)^{\theta}\right)
^\frac{1}{\alpha}.\end{aligned}
\label{cp:EE}
\end{equation}

\subsection{Technological side}

The second equation of our model can be obtained as the
{\it resource constraint}. 
In \cite{MendesM-2005} production is modeled using the Leontieff type production linear function, which determines  the amount of output $y_{t}$ given inputs $l_{t}$ (labor) and $k_{t-1}$ (capital): 
\begin{equation}
y_{t}=\min \left(  al_{t},bk_{t-1}\right)  .
\label{Leontieff_func}
\end{equation}
We use the another way.
We introduce the Cobb-Douglas production technology as a nonlinear function:
\begin{equation}
y_{t}=l_{t}^{\beta}k_{t-1}^{1-\beta} .
\label{Cobb-Douglas_func}
\end{equation}
The {\it profit} maximization function will give us the following {\it firms' problem}:
\begin{equation}
\begin{aligned}
\max \text{ }\underset{k_{t},l_{t}}{\Pi_{t}} = l_{t}^{\beta} k_{t-1}^{1-\beta} - w_t l_t - R_t k_{t-1}
,
\label{fp:max_obj_func_profit}
\end{aligned}
\end{equation}
which produces the following FOCs:%
\begin{equation}
\begin{aligned}
&\frac{\partial \Pi_{t}}{\partial l_t}  =0:  \beta \frac{y_t}{l_t} - w_t  =0,\\
&\frac{\partial \Pi_{t}}{\partial k_{t-1}}  =0: (1 - \beta) \frac{y_t}{k_{t-1}} - R_t  =0.\end{aligned}
\label{fp:FOCs}
\end{equation}
From equations \eqref{fp:FOCs}, we obtain the following
\begin{equation}
w_t = \beta \frac{y_t}{l_t} ,
\label{fp:FOC-1}
\end{equation}
\begin{equation}
R_t = (1 - \beta) \frac{y_t}{k_{t-1}} ,
\label{fp:FOC-2}
\end{equation}
therefore,
\begin{equation}
\frac{w_t}{R_t} = \frac{\beta}{(1 - \beta)} \frac{k_{t-1}}{l_t} .
\label{fp:ratio_CD_wt_Rt}
\end{equation}

Note that the production functions \eqref{Leontieff_func} and \eqref{Cobb-Douglas_func}
are thresholds of the CES function.
Now, we come back to the Leontieff production function \eqref{Leontieff_func}.
At time $t$ the inputs to production are labor of the time $t$ young, $l_{t},$
and the capital saved by the agents who were born at time $t-1$ and thus are
old at time $t$, $k_{t-1}$.
As the constant $a$ determines only the
coefficient of proportionality between the labor input and output,
it is normalized to 1.
The Leontieff production technology implies that the firm
maximizing its profits will use its inputs in a fixed proportion,
so that
\begin{equation}
y_{t}=l_{t}=bk_{t-1} ,
\label{Leo_func_lk}
\end{equation}
 therefore,
\begin{equation}
\frac{k_{t-1}}{l_t} = \frac{1}{b} .
\label{fp:ratio_Leontieff}
\end{equation}

Thus, we can say that relation between of the real wage rate $w_t$ and the
gross real interest rate $R_t$ is defined as the limit state of CES function
through the Cobb-Douglas production technology \eqref{fp:ratio_CD_wt_Rt}.
Hence, using the same ratio of $\frac{w_t}{R_t}$
as in the Cobb-Douglas production technology case \eqref{fp:ratio_CD_wt_Rt}
and $\frac{k_{t-1}}{l_t}$ ratio as in the Leontieff technology case \eqref{fp:ratio_Leontieff},
we get
\begin{equation}
\frac{R_t}{w_t} = b \frac{(1 - \beta)}{\beta} .
\label{fp:relation_wt_Rt}
\end{equation}
Moreover, we can use relations from Cobb-Douglas case \eqref{fp:FOC-2},
taking into account \eqref{Leo_func_lk} and \eqref{fp:relation_wt_Rt}, then
\begin{equation}
R_t = (1 - \beta)  \frac{y_t}{k_{t-1}} = (1 - \beta) \frac{b k_{t-1}}{k_{t-1}} = (1 - \beta) b ,
\label{Rt}
\end{equation}
and
\begin{equation}
w_t = \beta  \frac{y_t}{l_t} = \beta \frac{l_t}{l_t} = \beta.
\label{wt}
\end{equation}

In order to get the second equation of the OLG model, we then need to determine
how is the good produced at time $t,$ $y_{t}$, allocated.
The total amount of the good produced is split between consumption
of the young alive at $t$, $c_{t}^{t}$, savings of the young alive at $t,$ $k_{t}$,
consumption of the old alive at $t,$ $c_{t}^{t-1}$,
and the {\it government spending} $g_{t}$.
The government spending is financed through the proportional tax on the young,
and the amount collected equals $\tau_{t}w_{t}l_{t}$.
The government spending isn't used for anything productive
and doesn't contribute to the utility of
consumers, that's why it does not appear anywhere in the problem of young
consumers \eqref{cp:obj-function}.
The only purpose of the government spending $g_{t}$ in this
model is to provide a {\it control variable}.

We rectify the resource constraint in model \cite{MendesM-2005}, to take into account not only the consumption of the young in the current period, but also the consumption of the old in both periods (i.e., when they were young). Therefore, to obtain the second equation of our model,
we now could write {\it resource constraint} %
\begin{equation}
l_{t}    =y_{t}=k_{t}+c_{t}^{t}+c_{t}^{t-1}+g_{t}=bk_{t-1}.
\label{resource-constraint}%
\end{equation}
From \eqref{resource-constraint}, taking into account \eqref{Leo_func_lk},
we obtain the following
\begin{equation}
k_{t-1}   =y_{t-1}-c_{t-1}^{t-1}-c_{t-1}^{t-2}-g_{t-1},
\end{equation}
and
\begin{equation}
l_{t}   =b\left(  l_{t-1} - c_{t-1}^{t-1} - c_{t-1}^{t-2} - g_{t-1}\right).
\end{equation}
Moving forward one period,  we obtain the second  equation that characterizes
the dynamics of the OLG model
\begin{equation}
l_{t+1}   =b\left(  l_{t} - c_{t}^{t} - c_{t}^{t-1} - g_{t}\right).
\label{main-eq2}
\end{equation}
However, we need to eliminate $c_{t}^{t-1}$ from the equation,
because it is the consumption of old cohort.
Moving backward one period in the second equation from \eqref{cp:FOCs}
and taking into account~\eqref{Leo_func_lk} and~\eqref{Rt}, we obtain
\begin{equation}
c_{t}^{t-1}  = (1 - \beta) l_{t}.
\label{ct-old}
\end{equation}
Finally, plugging last relation into \eqref{main-eq2}, we get the OLG model with control function $g_t$
\begin{equation}
\begin{cases}
\begin{aligned}
&c_{t+1}^{t}=\left( l_{t}^{\gamma} - \left(  c_{t}^{t}\right)^{\theta}\right)
^{1 /\alpha} , \\
&l_{t+1}   =b\left( \beta  l_{t} - c_{t}^{t}  - g_{t}\right). \end{aligned}
\label{OLG-model}
\end{cases}
\end{equation}
Equations \eqref{OLG-model} represent a nonlinear dynamical model of
two equations in two variables, $l_{t}$ and $c_{t}^{t},$ with one
control variable $g_t$ included additively.
Thus, the government spending is
\begin{equation}
g_{t} = \tau_{t} w_{t} l_{t} = \tau_{t} \beta l_{t} .
\label{control}
\end{equation}
So the  government spending is a share of today's wage bills, and thus of today's output.
Note that the government spending in the previous period, $g_{t-1}>0,$ enter
with the negative sign in the equation for the current output $y_{t}=l_{t}.$
This is because past period's government spending forced the agents born at
time $t-1$ (current old) to reduce both their first period consumption,
$c_{t-1}^{t-1},$ and the amount of capital they bought with their savings,
$k_{t-1}.$ With less capital, the Leontieff technology forces the firms in period
$t$ to demand less labor and produce less output.

\subsection{Control}

From an economic point of view, interventions in the model aimed at controlling chaotic dynamics can only be carried out by way of  introducing a variable that could be intentionally controlled, for example, taxes, government spending, government consumption or investment, etc. Therefore, in our model, we introduce government spending (non-fundamental variable), which is used as a control variable.
Thus, we avoid the problem arising in the model in \cite{MendesM-2005} if labor elasticity $\gamma$ is chosen as the control parameter. The parameter $\gamma$ is fundamental and cannot be changed, therefore it is not suitable for control.

We would have a control function as
$g_t = K \left( l_t - l_{t-m}\right)$ (for periodic orbit with period $m$).
Due to $g_{t} = \tau_{t} \beta l_{t}$ we can consider the proportional taxes with rate
$\tau_{t}$ as $\tau_{t} = \frac{K}{\beta} \left( 1 - \frac{l_{t-m}}{l_t} \right)$.
One note of caution. As is stated previously, the government spending is
financed through taxes on the labor income of the young.
Therefore, it cannot be larger than $w_{t}l_{t}$ or smaller than $0$,
which is condition that needs to be checked during simulations.
Also, given that the technology is Leontieff, the wage rate $w_{t}$
and the interest rate $R_{t+1}$ cannot be determined form the optimal conditions of the firm,
as is usually done in the economic literature.
They are, to a large degree, arbitrary.
It might therefore be advisable to use
a Cobb-Douglas production function~\eqref{Cobb-Douglas_func}
instead, which will allow to to introduce wage and interest rate in a non-arbitrary manner, while complicating the equations.

\section{Dynamics of the OLG model}

\subsection{Analysis of the uncontrolled OLG model}

Our OLG model with respect to variables $(c_t,l_t)$\footnote{Further on we will use $c_t=c_{t}^{t}$ since there is only the superscript $t$ in \eqref{OLG-model}.} and control $g_t$
is described by the following two-dimensional map
$\varphi : \mathbb{R}^2 \to \mathbb{R}^2$, where:
\begin{equation}\label{eq:olg}
    \varphi(c_t,l_t) = \left((l_t^\gamma - c_t^\theta)^{1/\alpha}, \ b ( \beta l_t - c_t - g_t)\right),
\end{equation}
and $0 < \alpha, \theta < 1$, $0 < \beta \leq 1$, $\gamma \geq 1$, $b > 1$ are parameters.
This map generates the following discrete-time dynamical model 
\begin{equation}\label{eq:olg:ds}
    \begin{cases}
        c_{t+1} = (l_t^\gamma - c_t^\theta)^{1/\alpha},\\
        l_{t+1} = b (\beta l_t - c_t - g_t),
    \end{cases} \qquad
    t \in \mathbb{Z}_+,
\end{equation}
which describes complex behavior of agents in conditions of economic equilibrium – a situation when supply and demand in all markets are balanced.

First, we consider the case $g_t = 0$.
To calculate the equilibria, we must solve a nonlinear system defined by
\[
    \begin{cases}
        c_{t} = (l_t^\gamma - c_t^\theta)^{\,1 / \alpha},\\
        l_{t} = b (\beta l_t - c_t).
    \end{cases}
\]
There are two equilibria in this model:
the first one $E_1 = (0,0)$ is trivial and always locally unstable,
while the second one, $E_2 = (c_\star > 0,l_\star > 1)$,
cannot even be calculated explicitly.

For the sake of simplicity, let us consider a special case
$$
  \theta = \alpha
$$
and using the following change of variables $c_t := c_t^{\,1 / \alpha}, l_t: = l_t$
and parameter $\lambda = 1 / \alpha$ rewrite initial map~\eqref{eq:olg}
in the following form:
\begin{equation}\label{eq:olg:sc}
    \psi(c_t, \, l_t) = \left(l_t^\gamma - c_t, \ b (\beta l_t - c_t^\lambda)\right),
\end{equation}
with the following constraints:
\begin{equation}\label{eq:olg:sc:param_cond}
 \lambda > 1, \quad 0 < \beta < 1, \quad \gamma > 1, \quad b > 1.
\end{equation}
For the dynamical model, generated by map~\eqref{eq:olg:sc}, i.e.
\begin{equation}\label{eq:olg:sc:ds}
    \begin{cases}
        c_{t+1} = l_t^\gamma - c_t,\\
        l_{t+1} = b (\beta l_t - c_t^\lambda),
    \end{cases} \qquad
    t \in \mathbb{Z}_+,
\end{equation}
it is possible to define all two equilibria analytically:
\[
  E_1 = (0,0), \quad
  E_2 = \left(\frac{1}{2}\left(\exp\bigg[\tfrac{\lambda \ln2+\ln\tfrac{\beta b-1}{b}}{\lambda \gamma - 1}\bigg]\right)^\gamma,
  \exp\bigg[\tfrac{\lambda \ln2+\ln\tfrac{\beta b-1}{b}}{\lambda \gamma - 1}\bigg]\right).
\]

\begin{figure}[!ht]
 \centering
 \subfloat[$\lambda = 3$, $\beta = 1$, $\gamma = 1$, $b = 1.54$]{
    \label{fig::olg:sc:chaos:simple}
    \includegraphics[width=0.5\textwidth]{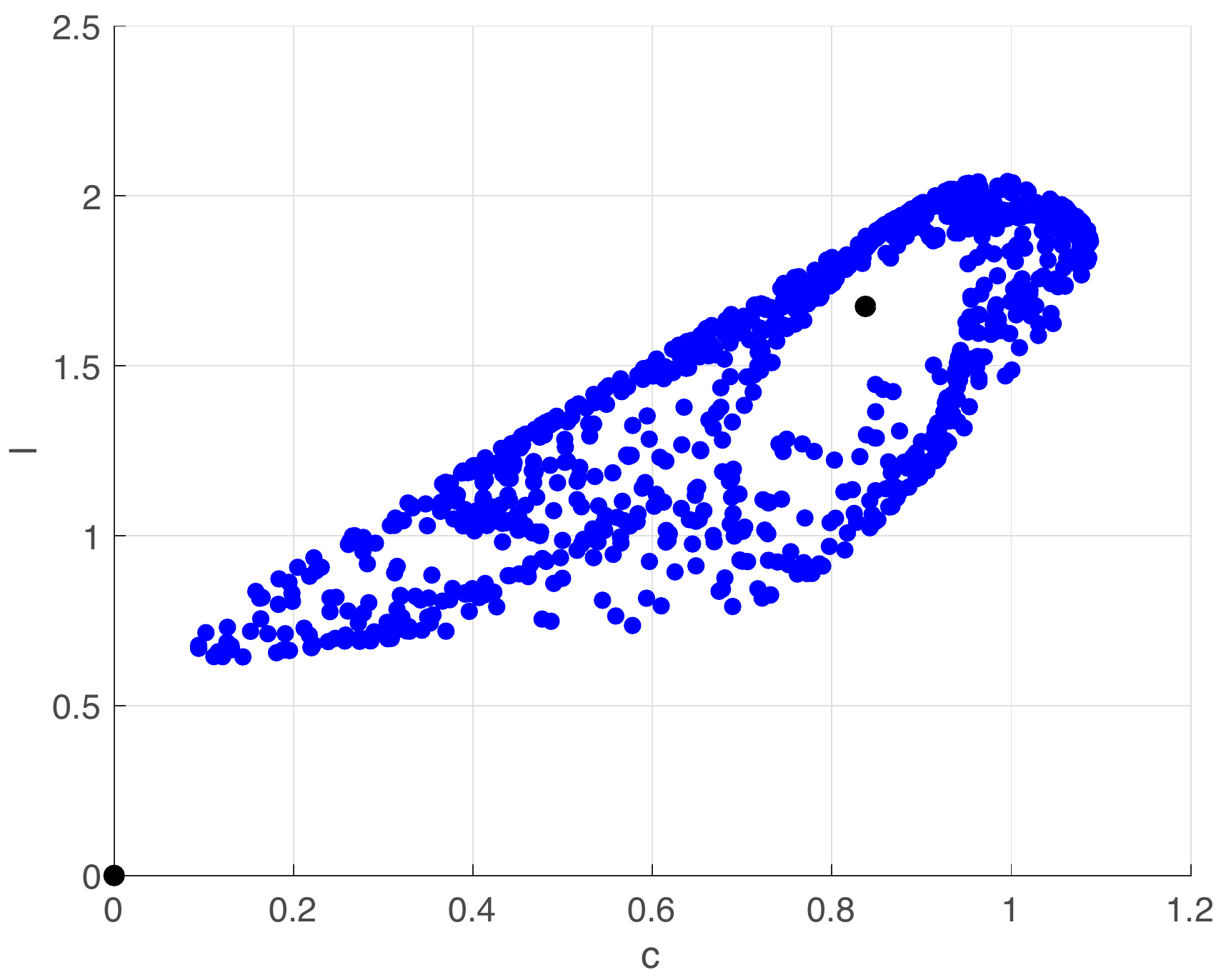}
  }~
 \subfloat[$\lambda = 3$, $\beta = 0.99$, $\gamma = 1.03$, $b = 1.54$]{
    \label{fig::olg:sc:chaos:economical}
    \includegraphics[width=0.5\textwidth]{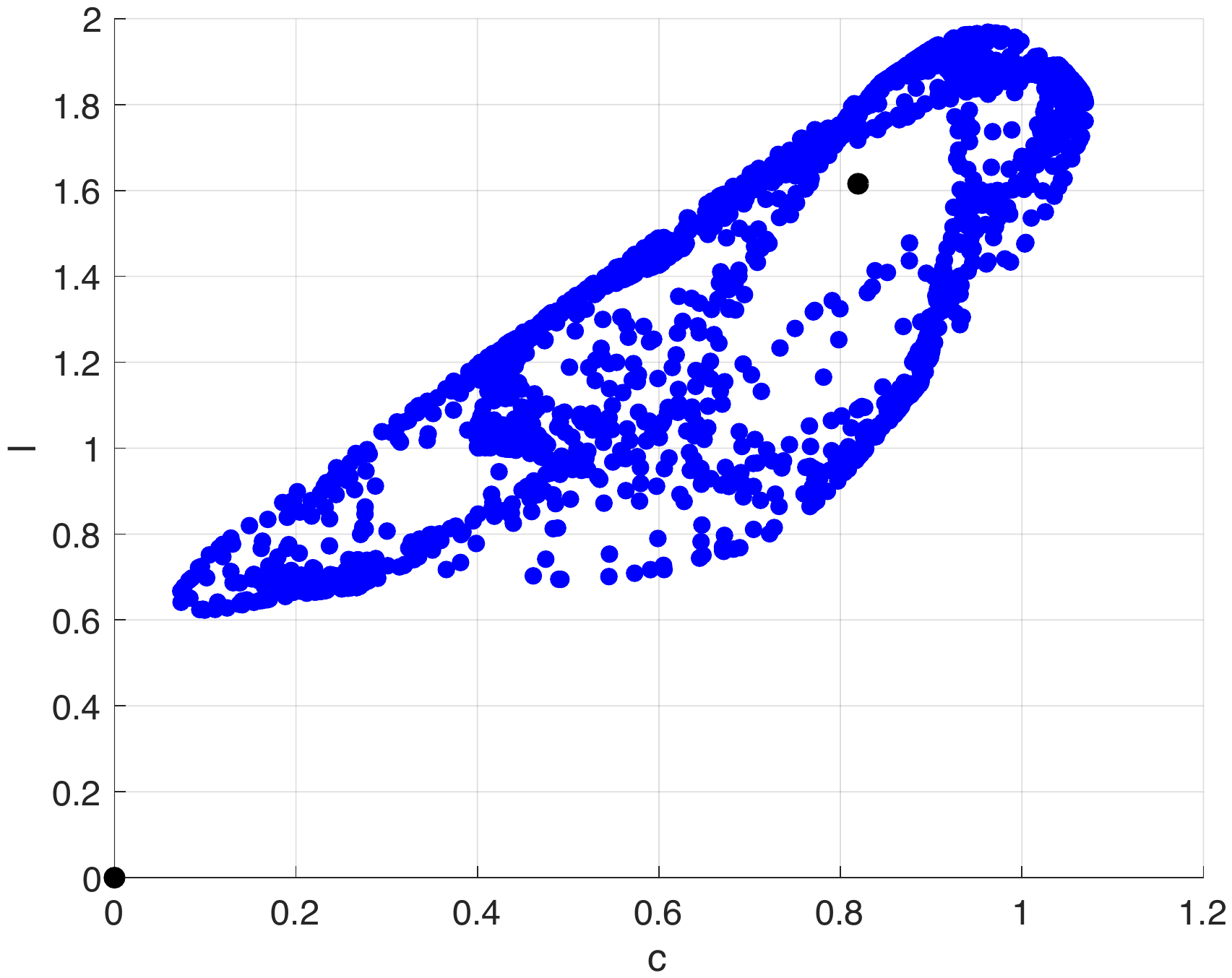}
  }
\caption{Chaotic attractors in OLG model  \eqref{eq:olg:sc} with parameters
\eqref{eq:olg:sc:param_simple}, \eqref{eq:olg:sc:param_economical}.}
\label{fig::olg:sc:chaos}
\end{figure}

In our work, for model~\eqref{eq:olg:sc} we study two groups of parameters
for which this model has chaotic behavior.
The first one,
\begin{equation}\label{eq:olg:sc:param_simple}
  \lambda = 3, \quad \beta = 1, \quad \gamma = 1, \quad b = 1.54,
\end{equation}
does not satisfy conditions \eqref{eq:olg:sc:param_cond},
but has only one nonlinearity with integer power, which makes
the study of~\eqref{eq:olg:sc}
much more simple.

The second one,
\begin{equation}\label{eq:olg:sc:param_economical}
  \lambda = 3, \quad \beta = 0.99, \quad \gamma = 1.03, \quad b = 1.54,
\end{equation}
satisfies conditions \eqref{eq:olg:sc:param_cond}, but
provides more issues, because of noninteger power involved in the first equation.

\subsection{Analysis of OLG model under control: searching for UPOs candidates}
\label{sec:search_upo}

Dynamics of agent's consumption and labor in models \eqref{eq:olg:ds} or \eqref{eq:olg:sc:ds} can be irregular with chaotic regime.
Agents and controller (for instance, a government or decision-makers) strive to suppress this undesirable phenomenon. To solve this problem one can either stabilize an unstable equilibrium (which is rather simple task),
or stabilize UPO embedded in a chaotic attractor. However, it could happen that for some {\it a~priori} given initial conditions, the agents are not capable of reaching the stationary state or even approaching it. In such a situation they would attempt to move to a trajectory with a more predictable dynamics, using a minimal control. The latter task requires to determine whether the corresponding chaotic attractor contains
any period-$m$ UPO,
i.e. whether for \eqref{eq:olg} or \eqref{eq:olg:sc}
there any solution for the corresponding equations:
\begin{equation}\label{eq:olg:upo_eq}
 (c_t, l_t) = \varphi^m(c_t, \, l_t), \quad \text{or} \quad
 (c_t, l_t) = \psi^m(c_t, \, l_t), \quad m = 1,2,\ldots.
\end{equation}
In order to find such control, we therefore assume that local attractor of the model is filled with periodic trajectories that are densely, and possibly uniformly, distributed in it. Then it is natural to find some periodic trajectories and under assumption of phase space mixing\footnote{A strict theoretical proof of mixing of trajectories and of ergodicity which involves constructing an ergodic measure is a complicated task beyond the scope of our study. Fundamental results obtained in this direction are presented, for example, in \cite{EckmannR-1985,BenedicksY-1993}.}. In our case, we could confirm existence of mixing in a reproducible numerical experiment, to expect that the current agents’ trajectory will over time enter neighborhood of the chosen periodic solution. Then one could “turn on” a minimal control intervention and switch the dynamics to this closest periodic trajectory. Similarly, a controller could use a control function (labor tax in our OLG model) to affect the expectations of agents and derive policy using calculations obtained with a mathematical model. In fact, the agents and the controller could act within paradigm of the same strategy and to solve similar tasks – forecasting and further control of dynamics aimed at selecting the periodic trajectory which corresponds to some feasible predictable solution. Here we face the following problem:
since the right-hand sides of models \eqref{eq:olg} and \eqref{eq:olg:sc}
represent polynomials with noninteger powers,
the search of periodic orbits, in general, can be made only numerically,
and as period of a periodic orbit increases, the search becomes more difficult,
and some point can't be performed even by using special functions for numerical
solving of nonlinear equations (e.g. {\ttfamily NSolve[\ldots]} in Wolframe Mathematica,
{\ttfamily vpasolve(\ldots)} in Matlab and {\ttfamily fsolve(\ldots)} in Maple).

In our work, to overcome this difficulty, as well as to further determine parameters of time-delay feedback control (DFC) within Pyragas procedure, we refer to the EAs. Today, there is a relatively rich set of EAs, which are divided into different subgroups according to the internal principles of their principles or the philosophy-natural processes from which they were derived (see, e.g. \cite{Zelinka-EAAI-2009}). The most well-known algorithms are, of course, genetic algorithms \cite{holland1992genetic}, \cite{mitchell1998introduction}, which represent classical EAs as well as DE \cite{price2013differential}, which is considered one of the most powerful EAs today \cite{Bilal-EAAI-2020}. Others are particle swarm \cite{eberhart1995particle} or SOMA \cite{davendra2016self}, which belong to the class of swarm algorithms. For more details on these fascinating algorithms, we recommend reading the literature \cite{dasgupta2013evolutionary}, \cite{back1993overview}.

We use three most powerful and commonly used optimization algorithms including DE and two versions of the SOMA as listed below:
\begin{itemize}
  \item \label{itm:EAs:DE} DE/rand/1/bin~\cite{StornP-1997}, with $NP = 50, Cr = 0.9, F = 0.7, MaxIter = 400$;
  \item \label{itm:EAs:SOMA} SOMA All To One strategy (SOMA ATO)~\cite{Zelinka-2004,Zelinka-2016}, with $popsize = 50, PathLength = 3.0, Step = 0.15, PRT = 0.33, MaxFEs = 20,000$;
  \item \label{itm:EAs:SOMAT3A} SOMA Team To Team Adaptive strategy (SOMA T3A)~\cite{Diep-2019,DiepZDS-2019}, with $popsize = 50, N_{jump} = 10, m = 10, n = 5, k = 10, MaxFEs = 20,000, PRT  = 0.05 + 0.90(FEs/MaxFEs),$ and $Step = 0.2 + 0.05\cos(4\pi FEs/MaxFEs)$.
\end{itemize}

In order to find the period-$m$ UPO we define the following cost function
\begin{equation}\label{eq:olg:upo:EA:cost_func}
{\rm CF}(c_t, l_t) = |(c_t, l_t) - \psi^m(c_t, \, l_t)|,
\end{equation}
and for $m = 2, \ldots 6$
will try to find its minimum over the bounded region
$$
  (c_t, l_t) \, \in \, [0.2, 1.2] \, \times \, [0.5, 2],
$$
where chaotic attractors, corresponding to parameters
\eqref{eq:olg:sc:param_simple}, \eqref{eq:olg:sc:param_economical},
are located (see Fig.~\ref{fig::olg:sc:chaos})

\begin{figure}[!ht]
 \centering
 \subfloat[$\alpha = 3$, $\beta = 1$, $\gamma = 1$, $b = 1.54$]{
    \label{fig::olg:sc:chaos:upo5:simple}
    \includegraphics[width=0.5\textwidth]{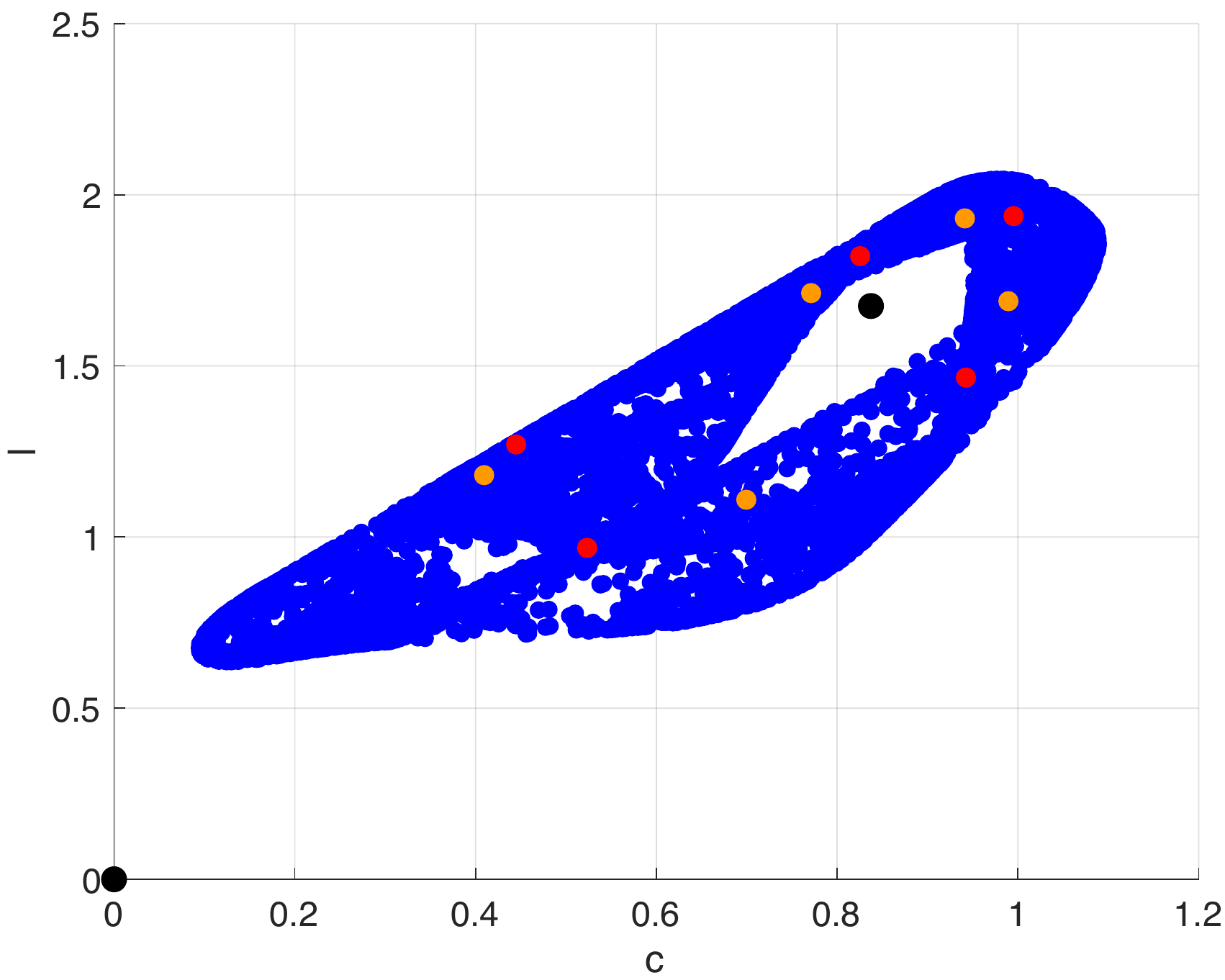}
  }~
 \subfloat[$\alpha = 3$, $\beta = 0.99$, $\gamma = 1.03$, $b = 1.54$]{
    \label{fig::olg:sc:chaos:upo5:economical}
    \includegraphics[width=0.5\textwidth]{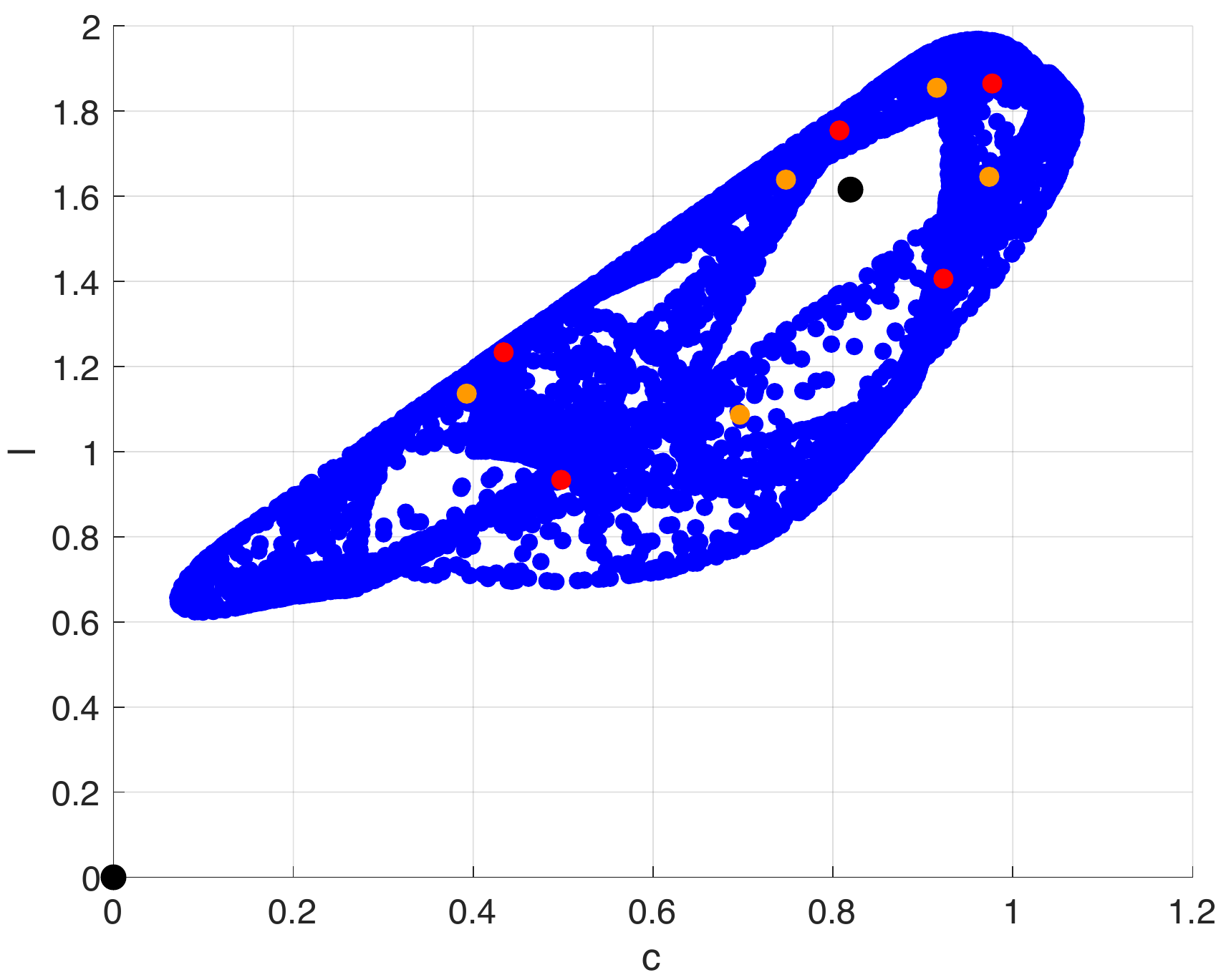}
  }
\caption{Period-5 UPOs (red, orange) embedded in the chaotic attractor (blue)
in the OLG model \eqref{eq:olg:sc} with parameters
\eqref{eq:olg:sc:param_simple}, \eqref{eq:olg:sc:param_economical}.}
\label{fig::olg:sc:chaos_upo5}
\end{figure}

As a result of this experiments we have the following conclusions:
\begin{enumerate}
  \item For OLG model \eqref{eq:olg:sc} with parameters \eqref{eq:olg:sc:param_simple}, \eqref{eq:olg:sc:param_economical} there are no periodic orbits with periods up to $m = 5$.

  \item The application of EAs
  allows us to find two different UPOs for $m = 5$
  (see Table~\ref{table:olg:upo5} and Fig.~\ref{fig::olg:sc:chaos_upo5})
  and also two UPOs for $m = 6$ (see Table~\ref{table:olg:upo6}).

\begin{table}[ht!]
\caption{Period-5 UPOs in the OLG model \eqref{eq:olg:sc}
with parameters~\eqref{eq:olg:sc:param_simple}, \eqref{eq:olg:sc:param_economical}.}
\label{table:olg:upo5}
\centering
\footnotesize
\begin{tabular}{
  |>{\centering}m{2cm}<{\centering}||
  >{\centering}m{3.3cm}<{\centering}|
  >{\centering}m{3.3cm}<{\centering}||
  >{\centering}m{3.3cm}<{\centering}|
  >{\centering}m{3.3cm}<{\centering}|}
  \hline
  \multirow{2}{*}{paremeters}
  & \multicolumn{2}{c||}{UPO \#1} & \multicolumn{2}{c|}{UPO \#2} \tabularnewline
  \cline{2-5}
  & $c_t$ & $l_t$ & $c_t$ & $l_t$ \tabularnewline
  \hhline{|=#=|=#=|=|}
  \multirow{5}{2cm}{
    $\lambda = 3$, $\beta = 1$,
    $\gamma = 1$, $b = 1.54$}
  & 0.9952568895406113 & 1.937721684767445 & 0.7712647366689733 & 1.7125896948743826 \tabularnewline
  & 0.9424647952268337 & 1.4659007926044583 & 0.9413249582054093 & 1.930857252566295 \tabularnewline
  & 0.5234359973776246 & 0.9682995881563105 & 0.9895322943608857 & 1.6890043947380275 \tabularnewline
  & 0.44486359077868604& 1.2703242075158732 & 0.6994721003771418 & 1.1089231076735928 \tabularnewline
  & 0.825460616737187 & 1.8207175062777983 & 0.409451007296451 & 1.1807157439654243 \tabularnewline
  \hhline{|=#=|=#=|=|}
  \multirow{5}{2cm}{
    $\lambda = 3$, $\beta = 0.99$,
    $\gamma = 1.03$, $b = 1.54$}
  & 0.49786228048149456 & 0.9336025046020485 & 0.9738430606936463 & 1.6456210369735524 \tabularnewline
  & 0.433817924634687 & 1.2333289032631394 & 0.6965540321966559 & 1.086625515820712 \tabularnewline
  & 0.807294938247322 & 1.7546020061352825 & 0.392783070128333 & 1.1362119035892562 \tabularnewline
  & 0.9771534043359568 & 1.8648192605089056 & 0.7477900042112287 & 1.63894775026562 \tabularnewline
  & 0.9228564283533333 & 1.4062615940651981& 0.9156305428565081 & 1.854778559891481 \tabularnewline
  \hline
  \end{tabular}
\end{table}

\begin{table}[ht!]
\caption{Period-6 UPOs in the OLG model \eqref{eq:olg:sc}
with parameters~\eqref{eq:olg:sc:param_simple}, \eqref{eq:olg:sc:param_economical}.}
\label{table:olg:upo6}
\centering
\small
\begin{tabular}{
  |>{\centering}m{2cm}<{\centering}||
  >{\centering}m{3.3cm}<{\centering}|
  >{\centering}m{3.3cm}<{\centering}||
  >{\centering}m{3.3cm}<{\centering}|
  >{\centering}m{3.3cm}<{\centering}|}
  \hline
  \multirow{2}{*}{paremeters}
  & \multicolumn{2}{c||}{UPO \#1} & \multicolumn{2}{c|}{UPO \#2} \tabularnewline
  \cline{2-5}
  & $c_t$ & $l_t$ & $c_t$ & $l_t$ \tabularnewline
  \hhline{|=#=|=#=|=|}
  \multirow{6}{2cm}{
    $\lambda = 3$, $\beta = 1$,
    $\gamma = 1$, $b = 1.54$}
  & 0.953087122876765 & 1.9918473398169698 & 0.2617410777499068 & 0.70299137243586226 \tabularnewline
  & 1.038760216940205 & 1.7341736162104449 & 0.4412502946859554 & 1.054992264689458 \tabularnewline
  & 0.6954133992702398 & 0.9445246126546086 & 0.6137419700035027 & 1.4923832432875062 \tabularnewline
  & 0.2491112133843688& 0.9366631654566522 & 0.8786412732840034 & 1.942247084462244 \tabularnewline
  & 0.6875519520722834 & 1.4186545006371223 & 1.0636058111782403 & 1.946447282891131 \tabularnewline
  & 0.7311025485648389 & 1.6841896714416038 & 0.8828414717128895 & 1.1445825494627964 \tabularnewline
  \hhline{|=#=|=#=|=|}
  \multirow{6}{2cm}{
    $\lambda = 3$, $\beta = 0.99$,
    $\gamma = 1.03$, $b = 1.54$}
  & 0.4273286812836567 & 1.0161364030089455 & 0.2360482546696026 & 0.9156617884384985 \tabularnewline
  & 0.58929581591653   & 1.429028575075392  & 0.6771964100935968 & 1.3757633891989363 \tabularnewline
  & 0.8551196850911698 & 1.8635444387760938 & 0.7117966011772947 & 1.619228199241378 \tabularnewline
  & 1.0435523163634506 & 1.8782149605648737 & 0.9310132570707518 & 1.9132975959437064 \tabularnewline
  & 0.8705169693966347 & 1.1134243718563632 & 1.0198912029882221 & 1.6742505107908452 \tabularnewline
  & 0.2465019913947314 & 0.6816233239954528 & 0.6804459848155441 & 0.9188249019130864 \tabularnewline
  \hline
  \end{tabular}
\end{table}
\end{enumerate}

In the next section,
let us apply Pyragas time-delay feedback control to stabilize these UPOs
and suppress chaos in system~\eqref{eq:olg:sc}, as well as discuss
the pros and cons of this approach and abilities of EA.

\subsection{Chaos suppression in the OLG model via DFC}

In order to apply DFC to model \eqref{eq:olg:sc:ds},
one needs to trace the values of the map at the previous iterations
to form a time-delayed feedback control.
This leads to the necessity of increasing the dimension of the initial map
by the artificial addition of equations defining iterations at the previous steps.
For instance, in order to stabilize the period-$m$ UPO in model~\eqref{eq:olg:sc:ds}
one needs to store the previous iterations of the coordinates up to $(c_{t-m},l_{t-m})$.
So, to apply DFC to this UPO by using additional control in the form~\eqref{control}
one needs to consider $m$ additional equations (to store only one coordinate $l_t$)
and the final system will has dimension $m + 2$;
if one needs a control involving components $(c_{t} - c_{t-5})$ and $(l_{t} - l_{t-5})$,
it will require consideration of $2 m$ additional equations
(to store both $c_t$ and $l_t$ variables),
and the final system will have dimension $2 m + 2$!

As we discussed above,
control is introduced into the model through the variable $g_t$,
which includes a proportional tax $\tau_t$ and labor $l_t$,
which allows us to implement the control in a natural way.
Consider OLG model~\eqref{eq:olg:sc:ds}
\[
    \begin{cases}
        c_{t+1} = l_t^\gamma - c_t,\\
        l_{t+1} = b (\beta l_t - c_t^\lambda - g_t),
    \end{cases} \qquad
    t \in \mathbb{Z}_+,
\]
with the DFC control in the form
\begin{equation}\label{eq:olg:dfc:control:economical}
  g_t = k_1 (l_t - l_{t-m}).
\end{equation}

Consider the following extended 7d map assuming this form of control:
\begin{equation}\label{eq:olg:dfc}
    \begin{cases}
        c_{t+1} = l_t^\gamma - c_t,\\
        l_{t+1} = b \big(\beta l_{t} - c_{t}^\alpha + k_1 \big(l_{t} - l^{(5)}_{t}\big)\big),\\
        l^{(1)}_{t+1} = l_t + k_2 (l_{t} - l^{(5)}_{t}), \\
        l^{(2)}_{t+1} = l^{(1)}_t + k_3 (c_{t} - c^5_{t}), \\
        l^{(3)}_{t+1} = l^{(2)}_t + k_4 (c_{t} - c^5_{t}), \\
        l^{(4)}_{t+1} = l^{(3)}_t + k_5 (c_{t} - c^5_{t}), \\
        l^{(5)}_{t+1} = l^{(4)}_t + k_6 (c_{t} - c^5_{t}). \\
    \end{cases}
\end{equation}
The corresponding Jacobi matrix of \eqref{eq:olg:dfc} has the following form:
\begin{equation}
  J(c_t,l_t) = \left(
    \begin{array}{ccccccc}
      -1 & \gamma l_t^{\gamma - 1} & 0 & 0 & 0 & 0 & 0 \\
      - \alpha b \, c_t^{\alpha-1} & b (\beta + k_1) & 0 & 0 & 0 & 0 & - b k_1 \\
      0 & 1 + k_2 & 0 & 0 & 0 & 0 & - k_2 \\
      0 & k_3 & 1 & 0 & 0 & 0 & -k_3 \\
      0 & k_4 & 0 & 1 & 0 & 0 & -k_4 \\
      0 & k_5 & 0 & 0 & 1 & 0 & -k_5 \\
      0 & k_6 & 0 & 0 & 0 & 1 & -k_6 \\
    \end{array}
  \right).
\end{equation}

According to DFC, our aim is to find such $k_1, \ldots, k_6$ for system \eqref{eq:olg:dfc}
to make initially unstable period-5 periodic orbits
from Table~\ref{table:olg:upo5} locally orbitally stable.
This is equivalent to have for the following fundamental matrix: 
\begin{equation}\label{eq:fund_mat}
  \Phi(c_t,l_t) = J(c_{t+4},l_{t+4}) \cdot J(c_{t+3},l_{t+3})
  \cdot J(c_{t+2},l_{t+2}) \cdot J(c_{t+1},l_{t+1}) \cdot J(c_t,l_t)
\end{equation}
calculated along the periodic orbits in Table \eqref{table:olg:upo5}
all absolute values of eigenvalues $\big\{\big|\lambda_i[\Phi]\big| \big\}_{i = 1}^7$ less than unity.
The latter is equivalent to have the spectral radius
(i.e. the largest absolute value of eigenvalues)
less than unity:
\[
  \rho[\Phi] =
  \max\big\{\big|\lambda_1[\Phi]\big|, \ldots, \big|\lambda_7[\Phi]\big| \big\} < 1.
\]

For simplicity, let us examine the controller with $k_3 = k_4 = k_5 = k_6 = 0$.
In order to find $k_1$, $k_2$ we define the cost function
\begin{equation}\label{eq:olg:dfc_k12:EA:cost_func1}
{\rm CF}(k_1, k_2) = \rho[\Phi]^2.
\end{equation}

The stabilization rate in the DFC depends on the stability of the periodic orbit
(more stable periodic orbit $\Rightarrow$ larger basin of attraction $\Rightarrow$ faster stabilization),
which in turn depends on how small it is possible to make absolute values for eigenvalues
of fundamental matrix \eqref{eq:fund_mat} by choosing the optimal values of the
gain coefficients $k_1$, $k_2$.
In order to adjust the 'stability rate' of the periodic orbit it is also possible to
experiment with the following cost functions, relying on the
arithmetic mean $\bar\lambda[\Phi] = \frac{1}{7}\sum_{i=1}^7 \big|\lambda_i[\Phi]\big|$:
\begin{equation}\label{eq:olg:dfc_k12:EA:cost_func2}
{\rm CF}(k_1, k_2) = \begin{cases}
  \rho[\Phi]^2, & \rho[\Phi] \geq 1\\
  {\rm CF_{temp}}(k_1, k_2), & \text{otherwise}
\end{cases}, \qquad
\text{where} \qquad
{\rm CF_{temp}}(k_1, k_2) = \bar\lambda[\Phi],
\end{equation}
\begin{equation}\label{eq:olg:dfc_k12:EA:cost_func3}
{\rm CF}(k_1, k_2) = \begin{cases}
  \rho[\Phi]^2, & \rho[\Phi] \geq 1\\
  {\rm CF_{temp}}(k_1, k_2), & \text{otherwise}
\end{cases}, \qquad
\text{where} \qquad
{\rm CF_{temp}}(k_1, k_2) = \sqrt{\frac{1}{7}\sum_{i=1}^7 \big(|\lambda_i[\Phi]| - \bar\lambda[\Phi] \big)^2}.
\end{equation}

Using EAs we try to find minimum of \eqref{eq:olg:dfc_k12:EA:cost_func1},
\eqref{eq:olg:dfc_k12:EA:cost_func2}, \eqref{eq:olg:dfc_k12:EA:cost_func3}
over the region
$$
  (k_1, k_2) \in (-1, 1) \times (-1, 1).
$$

As a result of this experiments we have the following conclusions:
\begin{enumerate}
  \item Using DFC with the 'economical' control \eqref{eq:olg:dfc:control:economical}
  it is possible to stabilize period-5 UPO\#1 (see Table~\ref{table:olg:upo5}) for
  OLG model~\eqref{eq:olg:sc:ds} with parameters \eqref{eq:olg:sc:param_simple}, \eqref{eq:olg:sc:param_economical}.
  All types of cost functions \eqref{eq:olg:dfc_k12:EA:cost_func1},
  \eqref{eq:olg:dfc_k12:EA:cost_func2}, \eqref{eq:olg:dfc_k12:EA:cost_func3} here works.

  \begin{table}[h!]
  \caption{Parameters $(k_1, k_2)$ for stabilization of UPO\#1 (see Table~\ref{table:olg:upo5})
  by DFC with the 'economical' control~\eqref{eq:olg:dfc:control:economical} in
  the OLG model \eqref{eq:olg:sc} with parameters~\eqref{eq:olg:sc:param_simple},
  \eqref{eq:olg:sc:param_economical}.}
  \label{table:olg:dfc:k12}
  \centering
  \small
  \begin{tabular}{
    |>{\centering}m{2.6cm}<{\centering}||
    >{\centering}m{3.3cm}<{\centering}|
    >{\centering}m{3.3cm}<{\centering}||
    >{\centering}m{2.8cm}<{\centering}|
    >{\centering}m{2.8cm}<{\centering}|}
    \hline
    \multirow{2}{*}{paremeters}
    & \multicolumn{2}{c||}{UPO \#1} &
    \multirow{2}{*}{$k_1$} & \multirow{2}{*}{$k_2$} \tabularnewline
    \cline{2-3}
    & $c_t$ & $l_t$ & & \tabularnewline
    \hhline{|=#=|=#=|=|}
    \multirow{5}{2.6cm}{
      $\lambda = 3$, $\beta = 1$,
      $\gamma = 1$, $b = 1.54$}
    & 0.9952568895406113 & 1.937721684767445 &
    \multirow{5}{2.8cm}{-0.13183153 -0.11874515} & \multirow{5}{2.8cm}{-0.88097316 -0.80453919} \tabularnewline
    & 0.9424647952268337 & 1.4659007926044583 & &  \tabularnewline
    & 0.5234359973776246 & 0.9682995881563105 & &  \tabularnewline
    & 0.44486359077868604& 1.2703242075158732 & &  \tabularnewline
    & 0.825460616737187 & 1.8207175062777983  & &  \tabularnewline
    \hhline{|=#=|=#=|=|}
    \multirow{5}{2.6cm}{
      $\lambda = 3$, $\beta = 0.99$,
      $\gamma = 1.03$, $b = 1.54$}
    & 0.49786228048149456 & 0.9336025046020485 &
    \multirow{5}{2.8cm}{-0.18808852 -0.13998896 -0.19513422 -0.14491652} &
    \multirow{5}{2.8cm}{-0.95598289 -0.88037581 -0.97834281 -0.94132827} \tabularnewline
    & 0.433817924634687 & 1.2333289032631394 & & \tabularnewline
    & 0.807294938247322 & 1.7546020061352825 & & \tabularnewline
    & 0.9771534043359568 & 1.8648192605089056 & & \tabularnewline
    & 0.9228564283533333 & 1.4062615940651981 & & \tabularnewline
    \hline
    \end{tabular}
  \end{table}

\item Using DFC with the 'economical' control \eqref{eq:olg:dfc:control:economical}
  we are not able to stabilize UPO\#2 (see Table~\ref{table:olg:upo5}) for
  OLG model \eqref{eq:olg:sc} with parameters \eqref{eq:olg:sc:param_simple},
  \eqref{eq:olg:sc:param_economical}.
  All cost functions \eqref{eq:olg:dfc_k12:EA:cost_func1},
  \eqref{eq:olg:dfc_k12:EA:cost_func2},
  \eqref{eq:olg:dfc_k12:EA:cost_func3} here does not work.

\end{enumerate}

\section{Optimal control problem to maximize the basin of attraction of stabilized UPO via DFC and EA}\label{sec:optimal_control_problem}

For given parameters of the model we can discover a periodic trajectory of the model and can show numerically that all solutions with initial conditions on a selected grid of points are always attracted to this periodic trajectory. However, implementation of numerical algorithms is complicated by numerical errors and the phenomenon of shadowing \cite{Pilyugin-2006}. Imagine that within this economic model the agents possess perfect foresight and target a true periodic trajectory, while the controller receives information from the mathematical model by running a numerical experiment and selects a pseudo-trajectory in a neighborhood of this periodic trajectory. In this case an informational disbalance between the agents’ and the controller’s actions is possible, because it becomes difficult to guess when the forecasted periodic trajectory gets sufficiently close and whether this is the true or the pseudo-trajectory. It is therefore obvious that “turning on” of the control function in order to move to the periodic orbit is a complicated problem. Control could fail and attraction to the target trajectory will not happen. Of course, one could start the numerical process again and obtain another realization, attempting to achieve successful control intervention. This, however, does not preclude global errors while making a control decision. Therefore, we could postulate a following problem. Using EA and Pyragas control methods, optimize control parameters in such a way that the basin of attraction is maximized. Then, even if the current dynamics is far from the target periodic trajectory, the control will work in a more flexible regime, and attraction to the target solution will nevertheless occur. This will allow to reduce probability of the control failing to stabilize the dynamics, and thus to reduce potentially catastrophic losses of the agents’ welfare.

In addition, using the Pyragas method allow to solve another important problem. Limited chaotic dynamics of the OLG model could also be observed when the model variables are outside of the feasible region of variables: for example, system's attractor is partially associated with negative values of some variables while the periodic solution is fully in the positive quadrant. In this case it is possible to use control for suppressing the chaotic behavior, return from the irregular to regular dynamics, and correct the behavior of the model in such a way that all variable values are located in the feasible set.

In this section, we demonstrate the abilities of EAs
to solve a more difficult control problem of
maximization of basins of attraction for stabilized UPOs.
For OLG model in form~\eqref{eq:olg:sc}, this problem is complicated
by the fact that this model is dichotomic and not dissipative
in the sense of Levinson (see e.g.~\cite{Levinson-1949,LeonovKM-2015-EPJST}),
thus, the well-known methods
of Lyapunov functions cannot be used to estimate
the basins of attraction.

Consider delayed map~\eqref{eq:olg:dfc}
with Pyragas control parameters
$(k_1, k_2) \in (-1, 1) \times (-1, 1)$, and
$k_3 = k_4 = k_5 = k_6 = 0$.
To compute the basin of attraction of stabilized UPO,
we specify a rectangular area
$$
  [c_{\min}, c_{\max}] \times [l_{\min}, l_{\max}], \quad \text{where} \quad
  c_{\min} = l_{\min} = 0, \ c_{\max} = 2, \ l_{\max} = 3
$$
and corresponding partition step
$$
  c_{\rm step} = l_{\rm step} = \varepsilon
$$
to generate a grid $\mathcal{B}_{\rm grid}(\varepsilon)$
consisting of $\big|\mathcal{B}_{\rm grid}(\varepsilon)\big| = (2/\varepsilon + 1) \times (3 / \varepsilon + 1)$
test points.
Here $|A|$ denotes the cardinality of a set $A$.

Grid points correspond to the first and second coordinates
of initial point for the OLG map~\eqref{eq:olg:dfc}:
$$
  (c_0, l_0) \in \mathcal{B}_{\rm grid}(\varepsilon).
$$

Five more (auxiliary) coordinates of the initial point
(which we had to introduce to be able to consider delayed map with lag = 5)
are considered as zeros:
$$
  l^{(1)}_0 = l^{(2)}_0 = l^{(3)}_0 = l^{(4)}_0 = l^{(5)}_0 = 0.
$$

For each initial point
$(c_0, l_0, 0, 0, 0, 0, 0)$ of map \eqref{eq:olg:dfc}
(where $ (c_0, l_0) \in \mathcal{B}_{\rm grid}(\varepsilon)$)
we will perform $N_{\rm iter}$ iterations to
trace the trajectory and depending on where this trajectory is attracted
to divide the set
\[
  \mathcal{B}_{\rm grid}(\varepsilon) =
  \mathcal{B}_{\rm inf}(\varepsilon, k_1, k_2) \, \cup \,
  \mathcal{B}_{\rm upo}(\varepsilon, N_{\rm iter}, k_1, k_2) \, \cup \,
  \mathcal{B}_{\rm other}(\varepsilon, N_{\rm iter}, k_1, k_2)
\]
into 3 subsets of initial points leading to different types of behavior:
\begin{itemize}
  \item {\bf Set $\mathcal{B}_{\rm inf}(\varepsilon, k_1, k_2)$}:
  points, from which system's trajectories go to infinity;
  \item {\bf Set $\mathcal{B}_{\rm upo}(\varepsilon, N_{\rm iter}, k_1, k_2)$}:
  points, from which system's trajectories after $N_{\rm iter}$ iterations
  tend to the stabilized UPO;
  \item {\bf Set $\mathcal{B}_{\rm other}(\varepsilon, N_{\rm iter}, k_1, k_2)$}:
  points, from which system's trajectories after $N_{\rm iter}$ iterations
  tend to some other attractors.
\end{itemize}
\smallskip
The aim of our study here is for fixed predefined $\varepsilon$ and $N_{\rm iter}$ to solve the following:\\
\smallskip
{\bf Optimal control problem}: to find such values of control parameters ($k_1$, $k_2$)
to maximize the number of points (cardinality) $|\mathcal{B}_{\rm upo}(\varepsilon, N_{\rm iter}, k_1, k_2)|$
of the basin of attraction $\mathcal{B}_{\rm upo}(\varepsilon, N_{\rm iter}, k_1, k_2)$ of the stabilized UPO:
\[
  \underset{(k_1, k_2)}{\operatorname{maximize}} \,
  \big|\mathcal{B}_{\rm upo}(\varepsilon, N_{\rm iter}, k_1, k_2)\big|
\]
w.r.t the chosen partition steps $c_{\rm step} = l_{\rm step} = \varepsilon$ and
grid points $\mathcal{B}_{\rm grid}(\varepsilon)$.

\bigskip
To solve this optimization problem using EAs
we can consider the following cost functions:
\begin{multline}\label{eq:olg:dfc_k12:EA:optimal:cost_func}
  {\rm CF}(k_1, k_2) = \big|\mathcal{B}_{\rm upo}(\varepsilon, N_{\rm iter}, k_1, k_2)\big|
  \, \to \, \max
  \qquad \text{or} \\
  {\rm CF}(k_1, k_2) =
  \big|\mathcal{B}_{\rm grid}(\varepsilon)\big|
  - \big|\mathcal{B}_{\rm upo}(\varepsilon, N_{\rm iter}, k_1, k_2)\big| \, \to \, \min.
\end{multline}

Solving such an optimization problem implies calculation
and examination of the behavior for the number of trajectories
of system~\eqref{eq:olg:dfc} equal to $|\mathcal{B}_{\rm grid}(\varepsilon)|$
at each iteration of the evolutionary algorithm.
As $\varepsilon$ decreasing (to consider more dense grid of points)
and $N_{\rm iter}$ increasing (to reveal the limiting behavior more accurately
and to cut-off long transient regimes), it becomes an extremely time and resource consuming
computational procedure.
In order to speed up this procedure, we implement it on two
powerful HPCs at IT4Innovations National Supercomputing Center of the Czech Republic.
We use Parallel Computing Toolbox in Matlab R2015b and R2018a 64-bit versions to implement on two powerful clusters of Barbora and Salomon at IT4Innovations National Supercomputing Center\footnote{https://www.it4i.cz/} of the Czech Republic, as listed below:

\begin{itemize}
  \item {\bf Barbora Cluster}~\cite{BarboraHPC-info}: under Red Hat Enterprise Linux 7.x operating system within 4 compute nodes, 36 cores of 2x18-core Intel Cascade Lake 6240 processors 2.6 GHz, at least 192 GB of RAM for each node (144 workers);
  \item {\bf Salomon Cluster}~\cite{SalomonHPC-info}: under CentOS 7.x Linux operating system within 20 compute nodes, 2x Intel Xeon E5-2680v3, 2.5 GHz, 2x12 cores for each node (480 workers), and 5.3 GB per core, DDR4@2133 MHz.
\end{itemize}
In total 624 workers with 144 hours of calculation time at each run.

\begin{table}[!ht]
\caption{For procedure parameters $\varepsilon = 0.01$, $N_{\rm iter} = 10000$,
  optimal parameters $(k_1, k_2)$ for stabilization of UPO\#1 by DFC with
  maximal basin of attraction for the OLG model~\eqref{eq:olg:sc}
  with parameters~\eqref{eq:olg:sc:param_simple}.}
\label{table:olg:dfc:k12:compBA}
\centering
\small
\begin{tabular}{
  |>{\centering}m{1.9cm}<{\centering}|
  >{\centering}m{2.3cm}<{\centering}||
  >{\centering}m{3.3cm}<{\centering}|
  >{\centering}m{3.3cm}<{\centering}|
  >{\centering}m{3.4cm}<{\centering}|}
  \hline
  \multirow{2}{*}{HPC} & \multirow{2}{*}{\makecell[cc]{Evolutionary \\ algorithm}}
  & \multicolumn{2}{c|}{Best member} &
  \multirow{2}{*}{$\big|\mathcal{B}_{\rm upo}(\varepsilon, N_{\rm iter}, k_1, k_2)\big|$}
  \tabularnewline
  \cline{3-4}
  & & $k_1$ & $k_2$ & \tabularnewline
  \hhline{|=|=#=|=|=|}
  \multirow{3}{*}{Barbora} & {\footnotesize DE/rand/1/bin} & -0.149784522963517 & -0.886566828622060 & 5299 \tabularnewline
  & {\footnotesize SOMA} & -0.146630592831770 & -0.873678217705145 & 5631 \tabularnewline
  & {\footnotesize SOMA T3A} & -0.141036797787322 & -0.856600665432177 & 4724 \tabularnewline
  \hhline{|=|=#=|=|=|}
  \multirow{3}{*}{Salomon}
  & {\footnotesize DE/rand/1/bin} & -0.124379557961065 & -0.873896102347582 & 2180 \tabularnewline
  & {\footnotesize SOMA} & -0.146540835618650 & -0.879197230586984 & 5693 \tabularnewline
  & {\footnotesize SOMA T3A} & -0.147295233453928 & -0.871914877265756 & 5728 \tabularnewline
  \hline
  \end{tabular}
\end{table}

\begin{figure}[!ht]
 \centering
 \noindent\makebox[\textwidth]{%
 \includegraphics[width=0.7\textwidth]{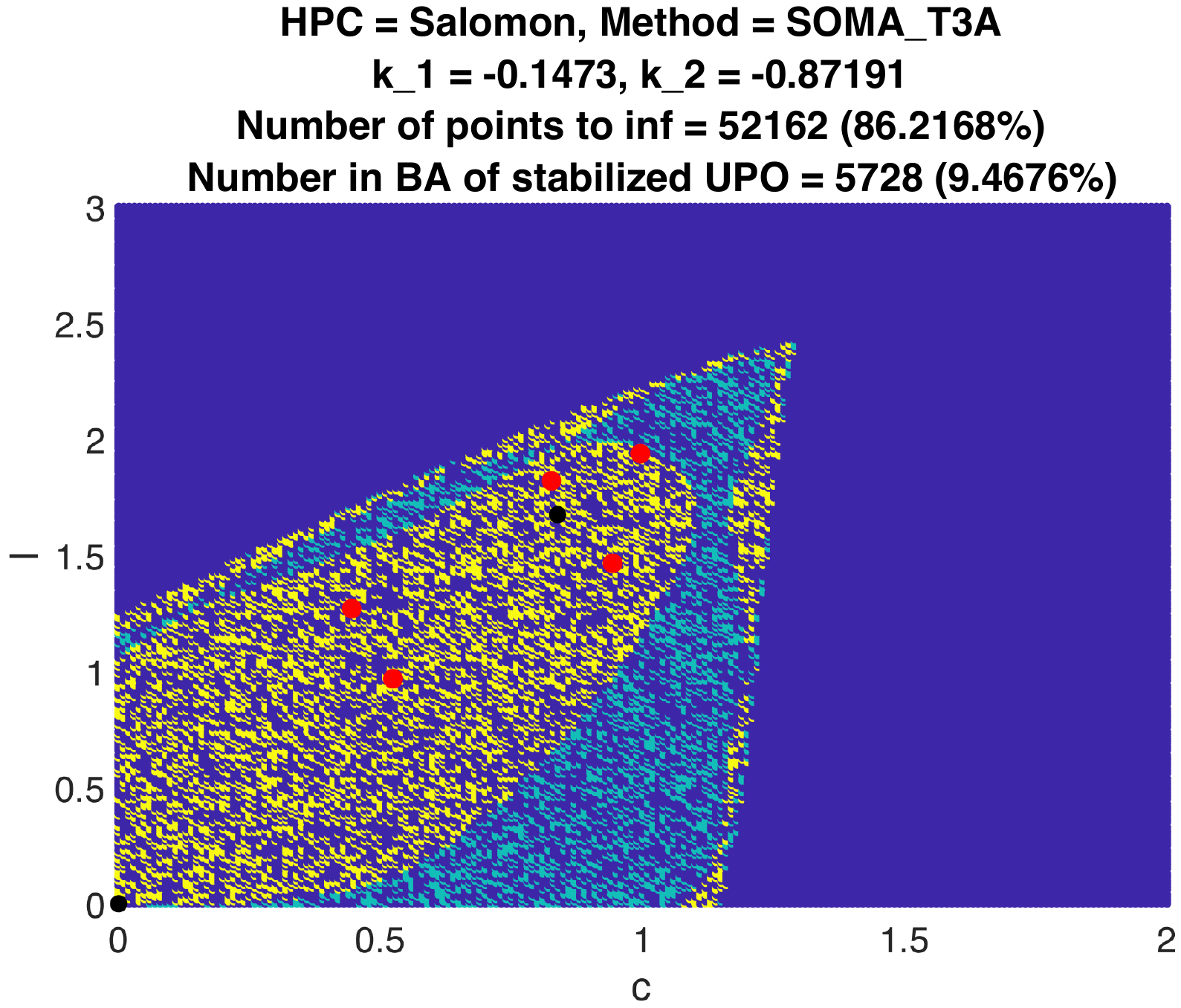}}
\caption{Basin of attraction (yellow) of stabilized UPO (red)
with respect to optimal control parameters $(k_1, k_2)$ for system~\eqref{eq:olg:dfc}
with parameters $\lambda = 3$, $\beta = 1$, $\gamma = 1$, $b = 1.54$.
Purple domain corresponds to the points, from which trajectories go to infinity;
cyan domain corresponds the points, from which trajectories go to other attractors.}
\label{fig::olg:basin:upo}
\end{figure}

The results of maximal basin of attraction computation
for stabilized UPO\#1 (see Table~\ref{table:olg:upo5} and Fig.~\ref{fig::olg:basin:upo})
using three evolutionary
algorithms (see Section~\ref{sec:search_upo})
for system~\eqref{eq:olg:dfc} with parameters
$\lambda = 3$, $\beta = 1$, $\gamma = 1$, $b = 1.54$ and for
procedure parameters $\varepsilon = 0.01$, $N_{\rm iter} = 10000$
are presented in Table~\ref{table:olg:dfc:k12:compBA}.
The best value (i.e. the maximum number of points in
$\big|\mathcal{B}_{\rm upo}(\varepsilon, N_{\rm iter}, k_1, k_2)\big| = 5728$)
gave SOMA T3A launched on Salomon Cluster;
however, it's interesting that SOMA T3A launched on Barbora Cluster
gave one of the worst results
(i.e. $\big|\mathcal{B}_{\rm upo}(\varepsilon, N_{\rm iter}, k_1, k_2)\big| = 4724$).

The overall simulation consumed about 344,714.12 core hours for the computation,
equivalent to more than 39.35 continuously working years of the single CPU at 2.5 GHz.

\section{Conclusion}

In this paper, we studied the properties of dynamics identifying regular
and irregular modes, including chaotic ones in a discrete-time OLG model
to improve forecasting its behavior.
We constructed the OLG model with a control function.
For the resulting model, we showed that in the absence of control,
both regular and irregular behavior (periodic and chaotic) can be observed in it.
Irregular behavior does not allow forecasting the limiting dynamics of the model,
and therefore, decision-makers do not have the ability to predict
and regulate the expectations of agents.
The limiting chaotic dynamics of the OLG model can be observed
when the range of admissible values of the model variables is violated.

To reveal the chaotic regime of the model's functioning, we used EA
and found periodic trajectories embedded into the attractor.
We applied the effective approach based on AI technologies
and methods for stabilizing unstable dynamics for suppressing chaos
to move from irregular to regular dynamics in the model,
to correct behavior of the model, returning the values of variables
to the admissible set by using small adjustments to the model parameters.
To refine the found periodic trajectories, we used the Pyragas method,
then we synthesized a control with two nonzero control parameters
and select these parameters using the EA in such a way
that the periodic trajectory becomes stable.
Moreover, we selected the controls in such a way that the basin of attraction
to the stable limiting dynamics is maximized.
The combination of EAs with the Pyragas method
significantly increased the efficiency of chaotic behavior control
and allowed us much faster and fine-tuning of the control parameters
to achieve the desired state of the model and the improvement
of its forecasting behavior.

\section*{Acknowledgment}

The following grants are acknowledged for the financial support provided for this research: St.Petersburg State University grant (Pure ID 75207094) [section~1],
Leading Scientific Schools of Russia (project NSh-4196.2022.1.1) [section~2],
the Russian Science Foundation project 22-11-00172 [section~4], the Ministry of Education, Youth and Sports of the Czech Republic through the e-INFRA CZ (ID:90140), Grant of SGS No. SP2022/22, VŠB-Technical University of Ostrava, Czech Republic.

\clearpage
\appendix

\section{Procedure implementing chaos suppression in the OLG model via DFC}

\begin{lstlisting}[frame=single]
% OLG map with delay
function out = olgMapD(x, alpha, beta, gamma, b, K)

    out = zeros(7,1);

    % Coordinates:
    % x(1) = c_i; x(2) = l_i; x(3) = c_{i-1};  ... x(7) = l_{i-5};

    out(1) = x(2)^gamma - x(1);
    out(2) = b * ( beta * x(2) - x(1)^alpha + K(1) * (x(2) - x(7)));
    out(3) = x(2) + K(2) * (x(2) - x(7));
    out(4) = x(3);
    out(5) = x(4);
    out(6) = x(5);
    out(7) = x(6);
end
\end{lstlisting}

\begin{lstlisting}[frame=single]
clearvars; clc;

numMapIter = 10000;
totalIter = numMapIter;

% Parameters:
alpha = 3; beta = 1; gamma = 1; b = 1.54;

% Equilibria:
S1 = [0, 0];

l_eq = exp((alpha*log(2) + log((beta * b - 1)/b)) / (alpha*gamma - 1));
c_eq = 0.5 * l_eq^gamma;

S2 = [c_eq, l_eq];

olgUPO = [.4448636000, 1.270324210; ...
          .8254646766, 1.820723642; ...
          .9952568454, 1.937723390; ...
          .9424648241, 1.465901543; ...
          .5234360180, .9682995990];

% alpha = 3; beta = 0.99; gamma = 1.03; b = 1.54;
% olgUPO =  [.49786228048149456,   .9336025046020485; ...
%            .433817924634687,  1.233328903263139;  ...
%            .807294938247322,  1.7546020061352825; ...
%            .9771534043359568,   1.8648192605089056; ...
%            .9228564283533333,   1.4062615940651981];

[olgUPO_period, ~] = size(olgUPO);

K_EA = [-0.13; -0.9];

currPoint = [0.0, 0.1, 0, 0, 0, 0, 0];

olgSolPyr = zeros(numMapIter, 7);

olgSolPyr(1,:) = currPoint;

numPeriodChunks = 2;

trajTail = zeros(numPeriodChunks * olgUPO_period, 2);

for iMapIter = 2 : numMapIter
    olgSolPyr(iMapIter, :) = olgMapD( currPoint, alpha, beta, gamma, b, K_EA)';

    currPoint = olgSolPyr(iMapIter, :);

%     currPoint = feval(olgMapD, currPoint);

    % tends to infty
    if currPoint(2) < 0 || abs(currPoint(1)) > 2
        totalIter = iMapIter;
        break;
    end
end

% Plot
figure; hold on;
% Trajectory with Pyragas stabilization (last 1000 iterations):
scatter(olgSolPyr(1:end, 1), olgSolPyr(1:end,2), 'filled','MarkerEdgeColor',[0, 0.6, 0], 'MarkerFaceColor', [0, 0.6, 0]);
% scatter(olgSolPyr(numMapIter-4 : numMapIter,1), olgSolPyr(numMapIter-4 : numMapIter,2), 'filled','MarkerEdgeColor',[0, 0.6, 0], 'MarkerFaceColor', [0, 0.6, 0]);
% UPO
scatter(olgUPO(:, 1), olgUPO(:,2), 'filled', 'SizeData', 50, 'MarkerEdgeColor','red', 'MarkerFaceColor', 'red');
% Equilibria:
plot(S1(1), S1(2), '.', 'markersize', 20, 'Color', 'black');
plot(S2(1), S2(2), '.', 'markersize', 20, 'Color', 'black');

hold off; grid on; axis on;
xlabel('c');
ylabel('l');
set(gca, 'FontSize', 14);

figure; hold on;
% Trajectory with Pyragas stabilization:
scatter(1 : totalIter, olgSolPyr(1 : totalIter, 1), 'filled','MarkerEdgeColor',[0, 0.6, 0], 'MarkerFaceColor', [0, 0.6, 0]);
hold off; grid on; axis on;
xlabel('t (iteration)');
ylabel('c');
set(gca, 'FontSize', 14);

figure; hold on;
% Trajectory with Pyragas stabilization:
scatter(1 : totalIter, olgSolPyr(1 : totalIter, 7), 'filled','MarkerEdgeColor',[0, 0.6, 0], 'MarkerFaceColor', [0, 0.6, 0]);
% scatter(numMapIter-1000 : numMapIter, olgSolPyr(numMapIter-1000 : numMapIter, 2), 'filled','MarkerEdgeColor',[0, 0.6, 0], 'MarkerFaceColor', [0, 0.6, 0]);
hold off; grid on; axis on;
xlabel('t (iteration)');
ylabel('c');
set(gca, 'FontSize', 14);
\end{lstlisting}

\section{Procedure of solving an optimal control problem to maximize the basin of attraction
of stabilized UPO via DFC and EA}

\subsection{Procedure, implementing calculation
of the cost function~(\ref{eq:olg:dfc_k12:EA:optimal:cost_func}).}

\begin{lstlisting}[frame=single]
function costValue = dfcParamCostFun(k1, k2, alpha, beta, gamma, b, gridPoints, olgUPO, numMapIter, numPeriodicChunks, periodicTol, vicTols, iTol)
    % Current parameters for Pyragas control
    % K_EA = [k1; k2];
    % Evaluation of the grid of points
    %vectFlags = computeBA_parallel(@(x) olgMapD(x, alpha, beta, gamma, b, K_EA), gridPoints, olgUPO, numMapIter, numPeriodicChunks, periodicTol, vicTols);
    %%%%%%%%%%%%%%%%%%%%%%%%%%%%%%%%%%%%%%%%%%%%%%%%%%%%%%%%%%%%%%%%%%%%%%%
    [numGridPoints, ~] = size(gridPoints);
    [olgUPO_period, ~] = size(olgUPO);
    numVicTols = length(vicTols);
    % By default, set flags = -1, as if all trajectories tend to infty
    vectFlags = -ones(numGridPoints, numVicTols);
    global time_parfor
    time_temp = toc;
    parfor iPoint = 1 : numGridPoints % parfor here
        currPoint = [gridPoints{iPoint}, 0, 0, 0, 0, 0];
        isToInf = 0;
        trajTail = zeros(numPeriodicChunks * olgUPO_period, 2);
        %%%%%%%%%%%%%%%%%%%%%%%%%%%%%%%%%%%%%%%%
        out = zeros(7,1);
        num_temp = numMapIter - numPeriodicChunks * olgUPO_period;
        %%%%%%%%%%%%%%%%%%%%%%%%%%%%%%%%%%%%%%%%
        for iMapIter = 2 : numMapIter
            % currPoint = feval(olgMapD, currPoint);
            %%%%%%%%%%%%%%%%%%%%%%%%%%%%%%%%%%%%%%%%%%%%%%%%%%%%%%%%%%%%%%%
            % x = currPoint;
            % out = zeros(7,1);
            % Coordinates:
            out(1) = currPoint(2)^gamma - currPoint(1);
            out(2) = b * (beta * currPoint(2) - currPoint(1)^alpha + k1 * (currPoint(2) - currPoint(7)));
            out(3) = currPoint(2) + k2 * (currPoint(2) - currPoint(7));
            out(4) = currPoint(3);
            out(5) = currPoint(4);
            out(6) = currPoint(5);
            out(7) = currPoint(6);
            currPoint = out;
            %%%%%%%%%%%%%%%%%%%%%%%%%%%%%%%%%%%%%%%%%%%%%%%%%%%%%%%%%%%%%%%
            % terminate evolution of trajectory if it tends to infty
            if abs(currPoint(1)) > 2
                isToInf = 1;
                break;
            end
            % cut the transient process and get
            % the "tail" of trajectory to compare with UPO
            if iMapIter > num_temp
                trajTail(iMapIter - num_temp, :) = currPoint(1:2);
            end
        end
        % if trajectory is not tending to infty and (almost) periodic, then
        % save results of comparison of trajectory and UPO
        %%%%%%%%%%%%%%%%%%%%%%%%%%%%%%%%%%%%%%%%%%%%%%%%%%%%%%%%%%%%%%%%%%%
        % checkIsPeriodicTol
        trajPeriodicChunks_x = reshape(trajTail(:,1), olgUPO_period, []);
        trajPeriodicChunks_y = reshape(trajTail(:,2), olgUPO_period, []);
        diff_x = abs(trajPeriodicChunks_x(:, 1:end-1) - trajPeriodicChunks_x(:,2:end));
        diff_y = abs(trajPeriodicChunks_y(:, 1:end-1) - trajPeriodicChunks_y(:,2:end));
        isPeriodic = all(diff_x < periodicTol) && all(diff_y < periodicTol);
        if ~isToInf && isPeriodic
            %%%%%%%%%%%%%%%%%%%%%%%%%%%%%%%%%%%%%%%%%%%%%%%%%%%%%%%%%%%%%%%
            %vectFlags(iPoint, :) = compareTrajUPO(trajTail(1+end-olgUPO_period:end, :), olgUPO, vicTols);
            %numVicTols = length(vicTols);
            vectFlags_temp2 = zeros(1, numVicTols);
            traj_x = sort(trajTail(1+end-olgUPO_period:end, 1));
            traj_y = sort(trajTail(1+end-olgUPO_period:end, 2));
            UPO_x = sort(olgUPO(:,1));
            UPO_y = sort(olgUPO(:,2));
            for iVicTol = 1 : numVicTols
                currVicTol = vicTols(iVicTol);
                % 1 (true) - if tending (stabilizing) to the target UPO (within tolerance);
                % 0 (false) - if tending somewhere else, or not tending at all;
                vectFlags_temp2(iVicTol) = ...
                    all(ismembertol(traj_x, UPO_x, currVicTol, 'ByRows', true)) && ...
                    all(ismembertol(traj_y, UPO_y, currVicTol, 'ByRows', true));
            end
            vectFlags(iPoint, :) = vectFlags_temp2;
        end
    end
    time_parfor = time_parfor + toc - time_temp;
    %%%%%%%%%%%%%%%%%%%%%%%%%%%%%%%%%%%%%%%%%%%%%%%%%%%%%%%%%%%%%%%%%%%%%%%
    % Number of points in the basin of attraction
    numPointsBA = sum(vectFlags == 1);
    %[numGridPoints, ~] = size(gridPoints);
    costValue = (numGridPoints - numPointsBA(iTol));
end
\end{lstlisting}

\subsection{Main scripts to run the procedure using DE/rand/1/bin.}

\begin{lstlisting}[frame=single]
clearvars; clc;
diary('Results/DiaryFile_DE_ran1bin_10k.txt');
disp('Method: DE rand1bin 10k')
disp('---------------------------------------------------')
% Define bounds of the grid:
cMin = 0; cMax = 2;
lMin = 0; lMax = 3;
% Test values of partion step
% cStep = 1; lStep = 1;
% ''Real'' values of partion step
cStep = 0.01; lStep = 0.01;
% Generate grid of points
[C, L] = meshgrid(cMin:cStep:cMax, lMin:lStep:lMax);
vectC = C(:); vectL = L(:);
gridPoints = num2cell([C(:), L(:)], 2);
% Set parameters of OLG map (with chaos):
alpha = 3;  beta = 1; gamma = 1; b = 1.54;
% Define correstponding UPO to stabilize:
olgUPO = [.4448636000, 1.270324210; ...
          .8254646766, 1.820723642; ...
          .9952568454, 1.937723390; ...
          .9424648241, 1.465901543; ...
          .5234360180, .9682995990];
% Set number of interations of delayed OLG map:
numMapIter = 1e3*10;
% Length of trajoctory's "tail" = number of chunks * period
numPeriodicChunks = 2;
% Tolerance to consider trajoctory's "tail" being periodic
periodicTol = 1e-5;
% Set the size of vicinity of UPO to check stabilization:
vicTols = 1e-1;
iTol = 1;
%%%%%%%%%%%%%%%%%%%%%%%%%%%%%%%%%%%%%%%%%%%%%%%%%%%%%%%%%%%%%%%%%%%%%%%%%%%
% Define parameters for DE:
popsize = 50;
Max_Gen = 400; % <=> Max_FEs = 20000
Cr    = 0.9;  F     = 0.7;
Xmin  = -0.3; Xmax  = 0;
Ymin  = -1.0; Ymax  = -0.7;
Dim     = 2;
CostFunction = @(pop) arrayfun(@(k1, k2) dfcParamCostFun(k1, k2, alpha,...
    beta, gamma,b , gridPoints, olgUPO, numMapIter, numPeriodicChunks,...
    periodicTol, vicTols, iTol), pop(:,1), pop(:,2));
%%%%%%%%%%%%%%%%%%%%%%%%%%%%%%%%%%%%%%%%%%%%%%%%%%%%%%%%%%%%%%%%%%%%%%%%%%%
filetxt = 'Results/Results_DE_ran1bin_10k.txt';
filemat = 'Results/DFC_DE_ran1bin_10k_in.mat';
fileID = fopen(filetxt, 'a');
fprintf(fileID,'FEs,best_cost,X,Y,time_flow,time_step,time_parfor\n');
fclose(fileID);
global time_parfor time_step
time_parfor = 0; time_step = 0;
tic
% --------------------- Create Initial Population -------------------------
pop           = Xmin + (Xmax-Xmin).*rand(popsize,1);
pop(:,2)      = Ymin + (Ymax-Ymin).*rand(popsize,1);
fit           = CostFunction(pop);
FEs           = popsize;
[best_fit,id] = min(fit);
best_val    = pop(id,:);
Store_and_Display(filetxt,FEs,best_fit,best_val);
for gen = 1 : Max_Gen-1
    % Combined Steps: Mutation+Crossover+Selection...to generate new pop
    FM_mui  = rand(popsize,Dim) < Cr;
    A1      = randperm(popsize);
    A2      = circshift(A1,1);
    A3      = circshift(A2,2);
    newpop  = pop(A1,:) + F*(pop(A2,:)-pop(A3,:));
    newpop  = pop.*not(FM_mui) + newpop.*FM_mui;
    % Check the boundary and replace individuals
    newpop(:,1)  = max(newpop(:,1),Xmin);
    newpop(:,1)  = min(newpop(:,1),Xmax);
    newpop(:,2)  = max(newpop(:,2),Ymin);
    newpop(:,2)  = min(newpop(:,2),Ymax);
    % Evaluate the new population
    newfit       = CostFunction(newpop);
    FEs          = FEs + popsize;
    % Update the new population
    idx          = newfit <= fit;
    fit(idx)     = newfit(idx);
    pop(idx,:)   = newpop(idx,:);
    % Update the Global Best
    [min_fit,id] = min(newfit);
    if min_fit   < best_fit
        best_fit = min_fit;
        best_val = newpop(id,:);
    end
    %====Store and Display Results ========================================
    try
        Store_and_Display(filetxt,FEs,best_fit,best_val);
        save(filemat)
    catch
        fprintf('Error while saving! Skipped this time. \n')
    end
end   % END Loop
disp('DONE!')
diary off
%%%%%%%%%%%%%%%%%%%%%%%%%%%%%%%%%%%%%%%%%%%%%%%%%%%%%%%%%%%%%%%%%%%%%%%%%%%
\end{lstlisting}

\subsection{Main scripts to run the procedure using SOMA.}

\begin{lstlisting}[frame=single]
clearvars; clc;
diary('Results/DiaryFile_SOMA_Classic_10k.txt');
disp('Method: SOMA Classic 10k')
disp('---------------------------------------------------')
% Define bounds of the grid:
cMin = 0; cMax = 2;
lMin = 0; lMax = 3;
% Test values of partion step
% cStep = 1; lStep = 1;
% ''Real'' values of partion step
cStep = 0.01; lStep = 0.01;
% Generate grid of points
[C, L] = meshgrid(cMin:cStep:cMax, lMin:lStep:lMax);
vectC = C(:); vectL = L(:);
gridPoints = num2cell([C(:), L(:)], 2);
% Set parameters of OLG map (with chaos):
alpha = 3;  beta = 1; gamma = 1; b = 1.54;
% Define correstponding UPO to stabilize:
olgUPO = [.4448636000, 1.270324210; ...
          .8254646766, 1.820723642; ...
          .9952568454, 1.937723390; ...
          .9424648241, 1.465901543; ...
          .5234360180, .9682995990];
% Set number of interations of delayed OLG map:
numMapIter = 1e3*10;
% Length of trajoctory's "tail" = number of chunks * period
numPeriodicChunks = 2;
% Tolerance to consider trajoctory's "tail" being periodic
periodicTol = 1e-5;
% Set the size of vicinity of UPO to check stabilization:
vicTols = 1e-1;
iTol = 1;
%%%%%%%%%%%%%%%%%%%%%%%%%%%%%%%%%%%%%%%%%%%%%%%%%%%%%%%%%%%%%%%%%%%%%%%%%%%
% -------------- Initial Parameters of SOMA -------------------------------
Step    = 0.15;   % Define the Step parameter
PRT     = 0.33;   % Define the PRT parameter
popsize = 50;     % Define the number individuals of the population
PathLen = 3;      % Define the PathLength parameter
Max_FEs = 20000;  % Define the stop condition
Xmin  = -0.3; Xmax  = 0;
Ymin  = -1.0; Ymax  = -0.7;
Dim     = 2;
CostFunction = @(pop) arrayfun(@(k1, k2) dfcParamCostFun(k1, k2, alpha,...
    beta, gamma,b , gridPoints, olgUPO, numMapIter, numPeriodicChunks,...
    periodicTol, vicTols, iTol), pop(:,1), pop(:,2));
%%%%%%%%%%%%%%%%%%%%%%%%%%%%%%%%%%%%%%%%%%%%%%%%%%%%%%%%%%%%%%%%%%%%%%%%%%%
filetxt = 'Results/Results_SOMA_Classic_10k.txt';
filemat = 'Results/DFC_SOMA_Classic_10k_in.mat';
fileID = fopen(filetxt, 'a');
fprintf(fileID,'FEs,best_cost,X,Y,time_flow,time_step,time_parfor\n');
fclose(fileID);
global time_parfor time_step
time_parfor = 0; time_step = 0;
tic
% --------------------- Create Initial Population -------------------------
pop           = Xmin + (Xmax-Xmin).*rand(popsize,1);
pop(:,2)      = Ymin + (Ymax-Ymin).*rand(popsize,1);
fit           = CostFunction(pop);
FEs           = popsize;
[best_fit,id] = min(fit);
best_val    = pop(id,:);
Store_and_Display(filetxt,FEs,best_fit,best_val);
while FEs < Max_FEs
  [~,idL] = min(fit);
  leader  = pop(idL,:);
  % ------------ movement of each individual ----------------------------
  for j = 1 : popsize
    if j ~= idL
            moving       = pop(j,:);
            nstep        = (Step:Step:PathLen)';
            PRTVector    = rand(length(nstep),Dim) < PRT;
            newpop       = moving+(leader-moving).*nstep.*PRTVector;
      %-- Check the boundary and replace the Individuals
            newpop(:,1)  = max(newpop(:,1),Xmin);
            newpop(:,1)  = min(newpop(:,1),Xmax);
            newpop(:,2)  = max(newpop(:,2),Ymin);
            newpop(:,2)  = min(newpop(:,2),Ymax);
      %----- Evaluate the offspring ---------------------------------
            newfit       = CostFunction(newpop);
            FEs          = FEs + length(nstep);
      %----- Choose the best offspring ------------------------------
      [min_fit,id] = min(newfit);
            %----- Update the best value ----------------------------------
            if  min_fit <= fit(j)
                pop(j,:) = newpop(id,:);
                fit(j)   = min_fit;
                if min_fit   < best_fit
                    best_fit = min_fit;
                    best_val = newpop(id,:);
                end
            end
            %====Store and Display Results ================================
            try
                Store_and_Display(filetxt,FEs,best_fit,best_val);
                save(filemat)
            catch
                fprintf('Error while saving! Skipped this time. \n')
            end
            %--------------------------------------------------------------
    end % END if j ~= idL
  end  % for j = 1 : popSize
end   % END Loop
disp('DONE!')
diary off
%%%%%%%%%%%%%%%%%%%%%%%%%%%%%%%%%%%%%%%%%%%%%%%%%%%%%%%%%%%%%%%%%%%%%%%%%%%
\end{lstlisting}

\subsection{Main scripts to run the procedure using SOMA T3A.}

\begin{lstlisting}[frame=single]
clearvars; clc;
diary('Results/DiaryFile_SOMA_T3A_10k.txt');
disp('Method: SOMA T3A 10k')
disp('---------------------------------------------------')
% Define bounds of the grid:
cMin = 0; cMax = 2;
lMin = 0; lMax = 3;
% Test values of partion step
% cStep = 1; lStep = 1;
% ''Real'' values of partion step
cStep = 0.01; lStep = 0.01;
% Generate grid of points
[C, L] = meshgrid(cMin:cStep:cMax, lMin:lStep:lMax);
vectC = C(:); vectL = L(:);
gridPoints = num2cell([C(:), L(:)], 2);
% Set parameters of OLG map (with chaos):
alpha = 3;  beta = 1; gamma = 1; b = 1.54;
% Define correstponding UPO to stabilize:
olgUPO = [.4448636000, 1.270324210; ...
          .8254646766, 1.820723642; ...
          .9952568454, 1.937723390; ...
          .9424648241, 1.465901543; ...
          .5234360180, .9682995990];
% Set number of interations of delayed OLG map:
numMapIter = 1e3*10;
% Length of trajoctory's "tail" = number of chunks * period
numPeriodicChunks = 2;
% Tolerance to consider trajoctory's "tail" being periodic
periodicTol = 1e-5;
% Set the size of vicinity of UPO to check stabilization:
vicTols = 1e-1;
iTol = 1;
%%%%%%%%%%%%%%%%%%%%%%%%%%%%%%%%%%%%%%%%%%%%%%%%%%%%%%%%%%%%%%%%%%%%%%%%%%%
% -------------- Initial Parameters of SOMA -------------------------------
popsize = 50;     % Define the number individuals of the population
N_jump  = 10;     % Define the PathLength parameter
Max_FEs = 20000;  % Define the stop condition
m       = 10;     % The parameter m
n       = 5;      % The parameter n
k       = 10;     % The parameter k
Xmin  = -0.3; Xmax  = 0;
Ymin  = -1.0; Ymax  = -0.7;
Dim     = 2;
CostFunction = @(pop) arrayfun(@(k1, k2) dfcParamCostFun(k1, k2, alpha,...
    beta, gamma,b , gridPoints, olgUPO, numMapIter, numPeriodicChunks,...
    periodicTol, vicTols, iTol), pop(:,1), pop(:,2));
%%%%%%%%%%%%%%%%%%%%%%%%%%%%%%%%%%%%%%%%%%%%%%%%%%%%%%%%%%%%%%%%%%%%%%%%%%%
filetxt = 'Results/Results_SOMA_T3A_10k.txt';
filemat = 'Results/DFC_SOMA_T3A_10k_in.mat';
fileID = fopen(filetxt, 'a');
fprintf(fileID,'FEs,best_cost,X,Y,time_flow,time_step,time_parfor\n');
fclose(fileID);
global time_parfor time_step
time_parfor = 0; time_step = 0;
tic
% --------------------- Create Initial Population -------------------------
pop           = Xmin + (Xmax-Xmin).*rand(popsize,1);
pop(:,2)      = Ymin + (Ymax-Ymin).*rand(popsize,1);
fit           = CostFunction(pop);
FEs           = popsize;
[best_fit,id] = min(fit);
best_val    = pop(id,:);
Store_and_Display(filetxt,FEs,best_fit,best_val);
while FEs+N_jump < Max_FEs
    % ------------ Migrant selection: m -----------------------------------
    M = randperm(popsize,m);
    [~,Im] = mink(fit(M),n);
  % ------------ movement of each individual ----------------------------
    for j  = 1 : n
        %------------ Update PRT and Step parameters ----------------------
        PRT  = 0.05 + 0.90*(FEs/Max_FEs);
        Step = 0.2 + 0.05*cos(4*pi*FEs/Max_FEs);
        Migrant = pop(M(Im(j)),:);
        %------------- Leader selection: k --------------------------------
        K = randperm(popsize,k);
        [~,Ik] = mink(fit(K),2);
        Leader  = pop(K(Ik(1)),:);
        if M(Im(j)) == K(Ik(2))
            Leader  = pop(K(Ik(2)),:);
        end
        %-------------- Moving process ------------------------------------
        nstep        = Step*(1:N_jump)';
        PRTVector    = rand(N_jump,Dim) < PRT;
        newpop       = Migrant+(Leader-Migrant).*nstep.*PRTVector;
        %----- Checking Boundary and Replaced Outsize Individuals ---------
        newpop(:,1)  = max(newpop(:,1),Xmin);
        newpop(:,1)  = min(newpop(:,1),Xmax);
        newpop(:,2)  = max(newpop(:,2),Ymin);
        newpop(:,2)  = min(newpop(:,2),Ymax);
        %----- SOMA Re-Evaluate Fitness Fuction ---------------------------
        newfit       = CostFunction(newpop);
        FEs          = FEs + N_jump;
        %----- SOMA Accepting: Place Best Individual to Population---------
        [min_fit,id] = min(newfit);
        if  min_fit <= fit(M(Im(j)))
            pop(M(Im(j)),:) = newpop(id,:);
            fit(M(Im(j)))   = min_fit;
            %----- SOMA Update Global_Leader ------------------------------
            if  min_fit  < best_fit
                best_fit = min_fit;
                best_val = newpop(id,:);
            end
        end
        %====Store and Display Results ====================================
        try
            Store_and_Display(filetxt,FEs,best_fit,best_val);
            save(filemat)
        catch
            fprintf('Error while saving! Skipped this time. \n')
        end
        %------------------------------------------------------------------
    end  % for j  = 1 : n
end   % while FEs+N_jump < Max_FEs
disp('DONE!')
diary off
%%%%%%%%%%%%%%%%%%%%%%%%%%%%%%%%%%%%%%%%%%%%%%%%%%%%%%%%%%%%%%%%%%%%%%%%%%%
\end{lstlisting}

\end{document}